\documentclass[aps,twocolumn,showpacs,preprintnumbers,nofootinbib,prd,superscriptaddress,groupedaddress,10pt]{revtex4-1}

\def\l@subsubsection#1#2{}
\def\l@subsubsubsection#1#2{}
\makeatother

\usepackage[utf8]{inputenc}
\usepackage{comment}
\usepackage{subfigure}
\usepackage{graphicx,amssymb,amsmath,amsthm,amsfonts,epsfig,epsf}
\usepackage[usenames]{color}
\usepackage{epstopdf}
\definecolor{darkred}{rgb}{0.5,0,0}
\usepackage{latexsym}
\usepackage{array}
\usepackage{afterpage}
\usepackage{bm}
\usepackage{dcolumn}
\usepackage[utf8]{inputenc}
\usepackage{rotating}
\usepackage{longtable}

\usepackage{enumerate}
\usepackage{tensor,multirow}
\usepackage{url}
\usepackage{placeins}
\usepackage[linktocpage]{hyperref}
\usepackage{float}
\usepackage{mwe}
\def\be{\begin{eqnarray}}
\def\ee{\end{eqnarray}}
\usepackage{soul}

\begin{document}

\title{Stability of the Cauchy horizon in accelerating black-hole spacetimes}

\author{Kyriakos Destounis}
\email{kyriakos.destounis@uni-tuebingen.de}
\affiliation{Theoretical Astrophysics, IAAT, University of T\"ubingen, Auf der Morgenstelle 14, 72076 T\"ubingen, Germany}

\author{Rodrigo D. B. Fontana}

\affiliation{Universidade Federal da Fronteira Sul, Campus Chapec\'o-SC
Rodovia SC 484 - Km 02, CEP 89815-899, Brasil}

\author{Filipe C. Mena}

\affiliation{Centro de An\'alise Matem\'atica, Geometria e Sistemas Din\^amicos, Instituto Superior T\'ecnico, Universidade de Lisboa, Avenida Rovisco Pais 1, 1049-001 Lisboa, Portugal}
\affiliation{Centro de Matem\'atica, Universidade do Minho, 4710-057 Braga, Portugal}
\date{\today}
\begin{abstract}
\noindent{The extendibility of spacetime and the existence of weak solutions to the Einstein field equations beyond Cauchy horizons, is a crucial ingredient to examine the limits of General Relativity. Strong Cosmic Censorship serves as a firewall for gravitation by demanding inextendibility of spacetime beyond the Cauchy horizon. 
For asymptotically flat spacetimes, the predominance of the blueshift instability and the subsequent formation of a mass-inflation singularity at the Cauchy horizon have, so far, substantiated the conjecture. Accelerating black holes, described by the $C-$metric, are exact solutions of the field equations without a cosmological constant, which possess an acceleration horizon with similar causal properties to the cosmological horizon of de Sitter spacetime. Here, by considering linear scalar field perturbations, we provide numerical evidence for the stability of the Cauchy horizon of charged accelerating black holes. In particular, we show that the stability of Cauchy horizons in accelerating charged black holes is connected to quasinormal modes, we discuss the regularity requirement for which weak solutions to the field equations exist at the Cauchy horizon and show that Strong Cosmic Censorship may be violated near extremality.}
\end{abstract}
\maketitle

\section{Introduction} One of the many fundamental questions concerning black holes (BHs) is their internal anatomy \cite{Dafermos:2003wr}. The inner structure of BHs, together with the nature of singularities \cite{Ori:2000fi} lying deep inside, is of paramount importance to understand the global uniqueness of solutions  to the Einstein field equations given suitable initial data, as well as the fate of infalling observers.

While, the issue of plunging observers into static and neutral BHs is quite clear \cite{Sbierski:2015nta}, their journey takes an unpredictable turn if the BH is charged and/or rotating. In those cases, the observer's journey seems to continue unaffected through the interior of the BH, eventually emerging into a region where neither the geometry nor the fate of the observer can be determined uniquely by initial data. The boundary of deterministic evolutions is called Cauchy horizon (CH) and marks the division between the region where General Relativity (GR) is able to forecast spacetime developments and the region where predictability of the field equations is lost.

Although Kerr and Reissner-Nordstr\"om (RN) BHs are known to have CHs \cite{Hawking:1973uf}, those are highly symmetric solutions which, in more realistic physical grounds, will eventually be perturbed by small time-dependent fluctuations. The relaxation of perturbed BHs leads to the emission of gravitational radiation in the form of damped sinusoid oscillations, described by quasinormal modes (QNMs) \cite{Kokkotas:1999bd,Berti:2009kk,Konoplya:2011qq}. 

The CH of asymptotically flat BHs is expected to be unstable under those perturbations, yielding a spacetime singularity which effectively seals off the tunnel to regions where the field equations cease to make sense \cite{Simpson:1973ua}. Any observer approaching the CH would  measure a divergent energy flux \cite{Hartle} which leads to the formation of a ``mass-inflation'' singularity \cite{Poisson_PRL,Poisson:1990eh}, due to the uneven competition between the power-law cutoff of perturbations in the exterior \cite{Price:1971fb,Leaver:1985ax,Dafermos:2003yw,Dafermos:2010hb,Dafermos:2014cua,Gundlach1,Gundlach2} and the exponential blueshift effect triggered at the CH. That this occurs generically is the essence of the Strong Cosmic Censorship (SCC) conjecture \cite{Penrose69,christodoulou1999instability,Dafermos:2012np}, which states that weak solutions of the field equations that arise from proper initial data are future inextendible beyond CHs. 
 
If a positive cosmological constant is included in those settings, then 
the exponential decay rate of perturbations in the exterior \cite{Hintz:2015jkj,Hintz:2016gwb,Hintz:2016jak}, which in turn is controlled by the dominant QNMs \cite{BonyHaefner,saBarreto,Dyatlov:2011jd,Dyatlov:2013hba}, may possibly counterbalance the blueshift amplification at the CH \cite{Costa:2014yha,Costa:2014zha,Costa:2014aia,Costa:2017tjc}. This leads to a weaker singularity, where the tidal deformations imposed on the observer there are bounded \cite{Ori:1991zz} and weak solutions to the field equations may exist. Then, to test SCC, it all comes down to the regularity of scalar field solutions at the CH \cite{Hintz:2016gwb,Costa:2017tjc,Dafermos:2017dbw} and the calculation of  $\beta\equiv-\inf\{\text{Im}(\omega)\}/\kappa_-$ \cite{Cardoso:2017soq,Hintz:2015jkj,CostaFranzen}, where the numerator captures the decay rate of the dominant QNM $\omega$ and $\kappa_-$ is the surface gravity of the CH, which governs the exponential perturbation growth there. For SCC to hold in these asymptotically de Sitter settings, $\beta<1/2$, which guarantees the breakdown of field equations at the CH \cite{Klainerman:2012wt}. If on the other hand $\beta>1/2$, then SCC may be violated.

The conclusions of very recent studies are the following: near-extremally-charged Reissner-Nordstr\"om-de Sitter (RNdS) BHs violate SCC with scalar \cite{Cardoso:2017soq,Cardoso:2018nvb,Zhang1,Dias:2018ufh}, Dirac \cite{Destounis:2018qnb,Zhang2} and gravitational perturbations \cite{Dias:2018etb}, while Kerr-de Sitter BHs do not \cite{Dias:2018ynt}. For a list of contemporary studies see \cite{Dafermos:2018tha,Luna:2018jfk,Rahman:2018oso,Liu:2019lon,Gwak:2018rba,Gim:2019rkl,Etesi:2019arr,Rahman:2019uwf,Guo:2019tjy,Dias:2019ery,Gan:2019jac,Destounis:2019omd,Liu:2019rbq,Zhang:2019nye,Chen:2019qbz,Mishra:2019ged,Gan:2019ibg,Hollands:2019whz,Rahman:2020guv,Mishra:2020gce,Emparan:2020rnp,casals2020glimpses}. 

Most studies, so far, have been performed under the assumption that BHs are static (or stationary) objects which do not ``move'' in space. On the contrary, many BHs are found in binary systems. The gravitational wave emission from those binaries leads to the increment of the BH velocities. Thus, these BHs seem to move and accelerate with respect to our own reference frame. 

A starting point to describe accelerating BHs is the $C-$metric \cite{Weyl} which is an axisymmetric exact solution of the field equations with a boost symmetry \cite{griffiths_podolsky_2009}. Although its geometrical properties are well known \cite{griffiths_podolsky_2009}, the physics of accelerating BHs has not been so well understood and, in part, this is because an appropriate framework to study their thermodynamics was lacking until recently \cite{Gregory-PRL-2016, GREGORY2019191,Anabalon:2018qfv,Anabalon:2018ydc}. Within GR, the $C-$metric has also been used to investigate radiation at infinity \cite{ashtekar1981, Podolsky-2003,Bicak-Winicour, Winicour}. However, it has been beyond classical GR that applications of the $C-$metric had most impact, e.g. in studies about the creation of BH pairs \cite{Hawking-Ross-PRL-1995, Hawking:1997ia}, the splitting of cosmic strings \cite{Earley-PRL-1995} and, most notably, in the construction of the black ring solution in five dimensions \cite{Emparan-Reall-PRL-2002}.

The charged version of the $C-$metric possesses a CH, similar to that of RN \cite{Hawking:1997ia}. The additional feature of the $C-$metric is the existence of an acceleration horizon; a hypersurface beyond which any event is unobservable to the accelerating BH's light cone. Although the $C-$metric which we consider here has no cosmological constant, its causal structure shares many similarities with RNdS spacetimes \cite{Hawking:1997ia,Griffiths:2006tk}. The reason is that the acceleration horizon essentially disconnects infinity from the region between the event and acceleration horizon, where observers reside, and the boundary conditions for the wave equation are significantly altered.

The investigation of the stability of the Cauchy horizon in accelerated BHs has been hampered by the fact that, until very recently, there were no studies about the decay of scalar perturbations on those backgrounds. This was partly due to the non-trivial geometry of accelerating BHs and their horizons which are not spherically symmetric. Note though the preliminary study \cite{Horowitz:1996yb}, where a simplified analysis was performed using null radiation.

A recent numerical approach has revealed that linear scalar perturbations decay exponentially on the charged $C-$metric and, most importantly, their decay between the event and acceleration horizon is controlled by the dominant QNMs \cite{Destounis:2020pjk}. This result is in agreement with the rigorous proof of stability and exponential decay of perturbations in asymptotically de Sitter black holes \cite{Dyatlov:2011jd,Hintz:2015jkj,Hintz:2016gwb,Hintz:2016jak}.  So, as in RNdS, the exponential decay of perturbations in the exterior of the accelerating BH may possibly counterbalance the exponential blueshift at the CH. This is in contrast with the behavior of perturbations on non-accelerating asymptotically flat BHs, which exhibit a power-law cutoff at late times.  
 
Unlike the case of Kerr-dS \cite{Dyatlov:2011jd}, there is no rigorous mathematical proof yet about the spectral gap and QNM dominance close to the horizons of the charged $C-$metric. However, the methods of \cite{Dyatlov:2011jd} for Kerr-dS are expected to apply in our case, as the wave operator involved is considerably simpler in the absence of rotation. 

Another ingredient that was missing in order to study SCC in the charged $C-$metric was the calculation of an eventual $\beta$ threshold beyond which the conjecture could be violated. We argue that a threshold indeed exists and is still given by $\beta>1/2$. Then, by computing $\beta$ we provide clear numerical evidence which indicate that SCC may be violated in near-extremally-charged accelerating BHs. To our knowledge, this is the first stability result of the Cauchy horizon in accelerating BH spacetimes.

\section{The charged $C-$metric in brief} Spacetimes based on the charged $C-$metric can be interpreted as representing axisymmetric electrically-charged BHs, accelerating along the axis of symmetry due to the presence of a cosmic string \cite{Griffiths:2006tk,griffiths_podolsky_2009,Wells}. Such BHs are described by the line element 
\begin{align}
\label{Cmetric}
ds^2=\frac{1}{\Omega^2}\left(-f(r)dt^2+\frac{dr^2}{f(r)}+\frac{r^2 d\theta^2}{P(\theta)} 
+P(\theta)r^2 \sin^2\theta d\varphi^2\right),
\end{align}
with $\Omega=1-\alpha r \cos\theta$ 
and
\begin{align}\label{lapse}
f(r)&=\left(1-\frac{2M}{r}+\frac{Q^2}{r^2}\right)(1-\alpha^2 r^2),\\
P(\theta)&=1-2\alpha M \cos\theta+\alpha^2 Q^2 \cos^2\theta,
\end{align}
where $M$, $Q$ and $\alpha$ are related to the BH mass, charge and acceleration, respectively. We observe that although there is no cosmological constant term in \eqref{lapse} the acceleration parameter $\alpha^2$ is dimensionally equivalent to a cosmological constant (in geometrized units). Therefore, one can expect that $\alpha^2$ may play the role of an effective cosmological constant in \eqref{Cmetric}. For a generalization of the $C-$metric to include an actual cosmological constant, see \cite{Plebanski:1976gy}.
The metric \eqref{Cmetric} asymptotes to the RN solution as $\alpha\rightarrow 0$ and to the $C-$metric as $Q\rightarrow 0$. There is a curvature singularity at $r=0$, while the roots of $f(r)$ determine the causal structure of spacetime. There exist three null hypersurfaces at 
\begin{equation}
r=r_\alpha:=\alpha^{-1}, \,\,\,\,\,\,\,\,\,\,\,\,r=r_{\pm}:=M\pm\sqrt{M^2-Q^2}, 
\end{equation}
called the acceleration horizon $r_\alpha$, event horizon $r_+$ and Cauchy horizon $r_-$, which must satisfy 
$r_-\leq r_+\leq r_\alpha,$ where $\alpha\leq 1/r_+$.
If $M=Q$, then the event horizon coincides with the CH and the BH is extremal. Each horizon has a surface gravity given by \cite{Griffiths:2006tk,Gregory:2017ogk}:
\begin{equation}
\kappa_i=\Big|\frac{f^\prime(r)}{2}\Big|_{r=r_i},\,\,\,\,\,\,\,\,\,i\in\{-,+,\alpha\}.
\end{equation}
Conical singularities occur on the axis at $\theta=0$ and $\theta=\pi$  designating the existence of deficit or excess angles. If we assume that $\varphi\in  [0,2\pi C )$, where $C=1/P(\pi)$, then we can remove the conical singularity at $\theta=\pi$ to end up with a deficit angle at $\theta=0$ (see \cite{Griffiths:2006tk,Destounis:2020pjk} for a detailed analysis). The metric \eqref{Cmetric} can, therefore, be understood as a RN-like BH accelerating along the axis $\theta=0$ 
\cite{Kinnersley_1970,Griffiths:2006tk}.

We can conformally rescale \eqref{Cmetric} as $d\tilde{s}^2=\Omega^2 ds^2$. The asymptotic structure of the charged $C-$metric  
has been analyzed in \cite{Kinnersley_1970,ashtekar1981,Hawking:1997ia,Griffiths:2006tk}. 


 
\section{Quasinormal modes of scalar fields in the charged $C-$metric}\label{sec3} 
The charged $C-$metric \eqref{Cmetric} is a solution to the vacuum Einstein-Maxwell field equations. Therefore, the Ricci curvature vanishes and the wave equation for a minimally coupled massless scalar field $\phi$ is equivalent to the conformally-invariant wave equation \cite{Hawking:1997ia}. The latter is invariant under 
$\tilde{g}_{\mu\nu}\rightarrow \Omega^2 g_{\mu\nu}$, $\tilde{\phi}\rightarrow\Omega^{-1}\phi$ and can be written as
\begin{equation}
\label{conf2}
\Box_{\tilde g} \tilde\phi-\frac{1}{6}\mathcal{\tilde{R}}\tilde{\phi}=0,
\end{equation}
where $\mathcal{\tilde{R}}$ is the conformally rescaled Ricci curvature \cite{Wald-book} and $\Box_{\tilde g}:=\tilde{g}_{\mu\nu}\nabla^{\mu}\nabla^\nu$. 
We can separate \eqref{conf2} by choosing an ansatz for the scalar field
\begin{equation}
\tilde{\phi}=e^{-i\omega t}e^{i m\varphi}\frac{\Phi(r)}{r}\chi(\theta),
\end{equation}
where $\omega$ is the QNM and $m=m_0 P(\pi)$ the azimuthal quantum number that guarantees the periodicity of $\varphi$, with $m_0>0$ integer \cite{Bini:2014kga}. Then, \eqref{conf2} becomes
\begin{eqnarray}
\label{final_radial}
\frac{d^2\Phi(r)}{dr^2_*}+(\omega^2-V_r)\Phi(r)&=0,\\
\label{final_polar}
\frac{d^2\chi(\theta)}{dz^2}-(m^2-V_\theta)\chi(\theta)&=0,
\end{eqnarray}
where $dr_*={dr}/{f(r)}$ and $dz={d\theta}/({P(\theta)\sin\theta})$ and
\begin{align}
\label{pot_r}
V_r&=f(r)\left(\frac{\lambda}{r^2}-\frac{f(r)}{3r^2}+\frac{f^\prime(r)}{3r}-\frac{f^{\prime\prime}(r)}{6}\right),\\\nonumber
V_\theta&=P(\theta)\left(\lambda \sin^2\theta-\frac{P(\theta)\sin^2\theta}{3}+\frac{\sin\theta\cos\theta P^\prime(\theta)}{2}\right.\\\label{pot_theta}&\left.+\frac{\sin^2\theta P^{\prime\prime}(\theta)}{6}\right),
\end{align}
with $\lambda$ the separation constant. The existence and nature of the acceleration horizon forces us to restrict our evolution to the range $r_+<r<r_\alpha$, where \eqref{lapse} is positive and \eqref{Cmetric} has fixed signature, implying that $P(\theta)$ is positive for $\theta\in [0,\pi]$. Thus, for QNMs, the boundary conditions are divided in two categories:
\begin{align}
\label{bcs_radial}
\Phi(r\rightarrow r_+)\sim e^{-i\omega r_* }&,\,\,\,\,\,\Phi(r\rightarrow r_\alpha)\sim e^{i\omega r_* }\\\label{bcs_polar}
\chi(\theta\rightarrow 0)\sim e^{mz}&,\,\,\,\,\,\chi(\theta\rightarrow \pi)\sim e^{-mz}.
\end{align}
Conditions \eqref{bcs_radial} impose purely ingoing (outgoing) waves at the event (acceleration) horizon, 
while conditions \eqref{bcs_polar} are taken so that the scalar field is bounded at the interval boundaries of $\theta$ \cite{Hawking:1997ia}. By solving \eqref{final_polar}, subject to the boundary conditions \eqref{bcs_polar}, one obtains the eigenvalues $\lambda$, for a given $m_0$. Then, those eigenvalues can be substituted in \eqref{final_radial}, subject to the boundary conditions \eqref{bcs_radial}, to obtain a discrete set of QNMs $\omega$.

In \cite{Dyatlov:2011jd} it was shown that asymptotic solutions to the wave equation in the Kerr-dS metric can be expressed as a sum of QNMs plus other sub-dominant terms. This result provides a mathematical proof for the physical interpretation of QNMs as complex frequencies of (linear scalar) gravitational waves in Kerr-dS. However, there is no such result for the $C-$metric yet. Instead, in \cite{Destounis:2020pjk}, we provided strong numerical evidence that this is the case, i.e. that QNMs indeed dominate the asymptotic dynamics of solutions to the wave equation. 

Specifically, the results in \cite{Destounis:2020pjk} indicate that the late-time decay of scalar field perturbations in the charged $C-$metric follows the numerically extracted exponential law \cite{Destounis:2020pjk}
\begin{equation}
\label{expo}
{\phi}\sim e^{-\gamma t},\,\,\,\,\,\,\,\text{for}\,\,\,\,\,\,t\rightarrow\infty,
\end{equation}
where
$\gamma:=-\text{inf}_{mn}\{\text{Im}(\omega)\}$
is the smallest (in absolute value) imaginary part of all families of QNMs. This justifies the use of QNMs to study the decay of scalar perturbations in the $C-$metric and, ultimately, the regularity of the spacetime extensions using, as a proxy, the regularity requirements for the solutions $\phi$ to the wave equation \cite{Hintz:2015jkj}.

Although ${\phi}$ and ${g}_{\mu\nu}$ may not necessarily be continuously differentiable, heuristically one can still make sense of the Einstein-Maxwell field equations by multiplying with a smooth, compactly supported, test function $\psi$ and integrating in a neighborhood $\mathcal{V}$ around the CH. If the result is finite, we may consider weak solutions to the field equations. The energy momentum tensor of the conformally-coupled scalar field \cite{Charmousis_2009} leads, after integration, to the regularity requirement that ${\phi}$ should belong to the (Sobolev) function space $H^1_\text{loc}$ for weak solutions to exist at the CH.

By considering \eqref{final_radial} as $r\rightarrow r_-$, we find two independent mode solutions
\begin{align}
{\phi}_1\sim e^{-i\omega u},\,\,\,\,\,\,\,\,\,\,\,
{\phi}_2\sim e^{-i\omega u}|r-r_-|^{i\omega/\kappa_-},
\end{align}
where we have dropped the angular dependence and used outgoing null coordinates $u=t-r_*$, which are regular at the CH. There, ${\phi}_1$ is smooth, while ${\phi}_2$ is not necessarily so. Therefore, considering $\phi\in H^{1}_\text{loc}$, we require finiteness of $\int_\mathcal{V}(\partial_r\phi_2)^2 dr$ and arrive at the condition
\begin{equation}
\label{def-beta}
\beta:=\gamma/\kappa_->1/2,
\end{equation}
which is identical to the one derived rigorously in \cite{Hintz:2015jkj} for Kerr-dS.
Since the metric should share similar regularity requirements as $\phi$ \cite{Hintz:2016gwb,Costa:2017tjc,Dafermos:2017dbw}, then, for $\beta>1/2$, the corresponding BH spacetime should extend beyond the CH with metric in $H^1_\text{loc}$.

To obtain the separation constant $\lambda$ from \eqref{final_polar} for given $m_0$, we use the {\it Mathematica} package {\it QNMSpectral} developed in \cite{Jansen:2017oag} (based on pseudospectral collocation methods \cite{Dias:2015nua}) and confirm the validity of our results with the Frobenious method \cite{Destounis:2020pjk}.  
To calculate $\omega$, we use {\it QNMSpectral}, and confirm our results with the numerical scheme developed in \cite{Gundlach1}, based on the time-domain integration of \eqref{final_radial} and the application of the Prony method \cite{Berti:2007dg} to extract the QNMs. For an extensive QNM analysis of the charged $C-$metric, see \cite{Destounis:2020pjk}.

\section{Quasinormal modes and Cauchy horizon stability}
\begin{figure}[t]
\includegraphics[scale=0.14]{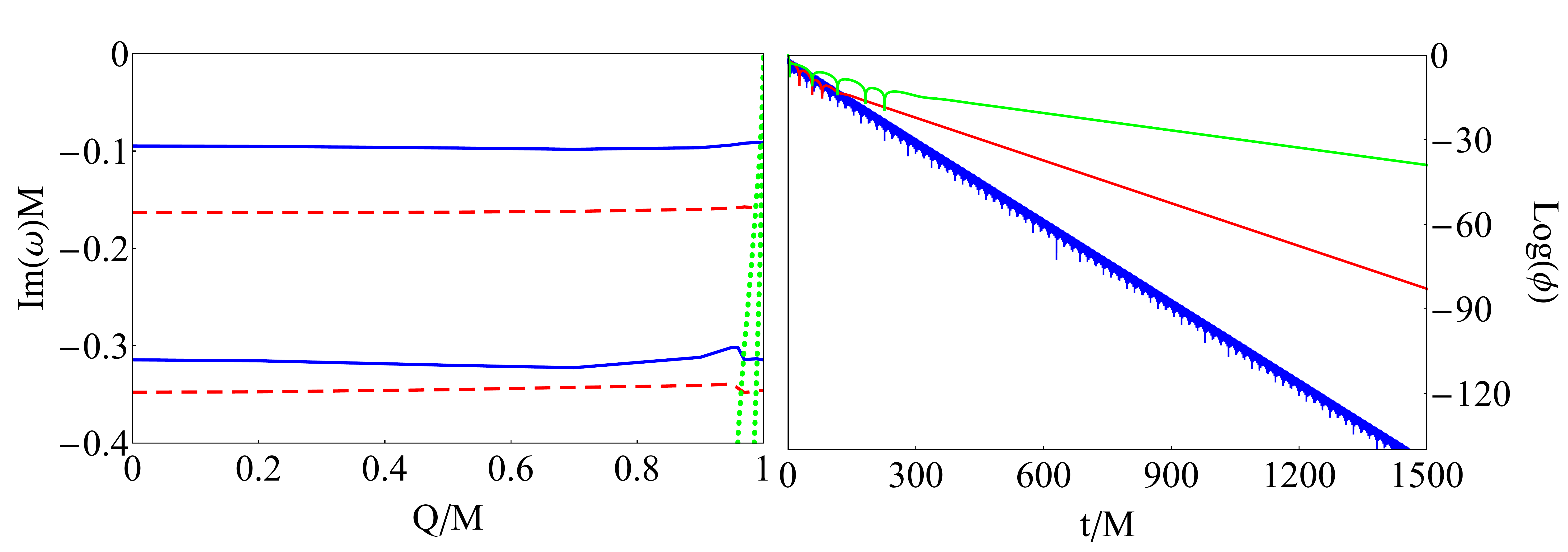}
\caption{Left: Imaginary parts of the lowest-lying quasinormal modes for $m_0=0$ and $\alpha M=0.15$ as a function of $Q/M$. The solid blue curves correspond to photon surface modes, the red dashed curves to acceleration modes and the green dotted curves to near extremal modes. Right: Time evolution of scalar perturbations on the charged $C-$metric with $m_0=10$, $Q/M=0.3$, $\alpha M=0.1$ (blue curve), $m_0=0$, $Q/M=0.3$, $\alpha M=0.05$ (red curve) and $m_0=0$, $Q/M=0.9995$, $\alpha M=0.6$ (green curve). The dominant mode extracted from the blue, red and green curves, at late times, belong to the photon surface, acceleration and near extremal family, respectively.}
\label{QNM}
\end{figure}

\begin{figure}[t]
\includegraphics[scale=0.145]{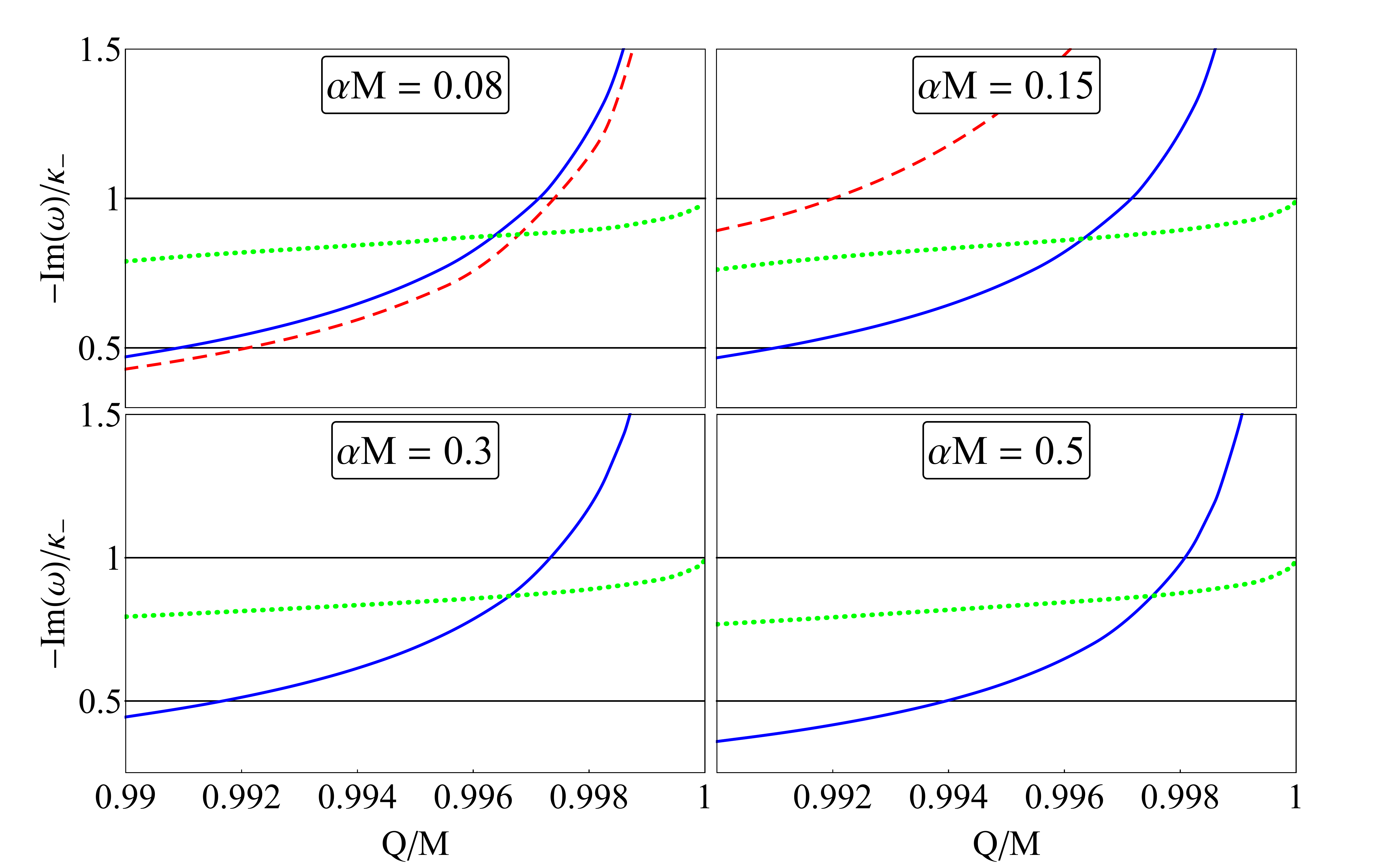}
\caption{The parameter $\beta$ calculated from the dominant quasinormal modes of a conformally-coupled scalar field propagating on the fixed near-extremally-charged $C-$metric, for various acceleration parameters $\alpha M$. The black horizontal lines denote $\beta=1/2$ (lower) and $\beta=1$ (upper), respectively. The blue curves correspond to the $m_0=10$ approximation of the dominant photon surface modes, the red, dashed curves correspond to the $m_0=0$ dominant acceleration modes, while the green dotted correspond to the $m_0=0$ dominant near extremal modes.}
\label{beta}
\end{figure}

Our numerics indicate the existence of three families of QNMs which antagonize each other in different regions of the parameter space, as show in Fig. \ref{QNM} and \ref{beta}. 

The first family of QNMs consists of the usual oscillatory modes obtained with standard WKB tools. We refer to them as ``photon surface'' (PS) modes (in blue in Fig. \ref{QNM} and \ref{beta}). These modes asymptote the oscillatory QNMs of non-accelerating BHs when $\alpha\rightarrow 0$. The dominant mode of the PS family is obtained at the large $m$ limit. Our numerics indicate that $m_0=10$ gives a very good approximation of the actual dominant mode \cite{Destounis:2020pjk}.

The second family of modes, the acceleration QNMs (in red in Fig. \ref{QNM} and \ref{beta}), is a novel family first found in \cite{Destounis:2020pjk}, which depends linearly on the acceleration parameter and vanishes when $\alpha\to 0$. The acceleration modes are purely imaginary and arise due to the presence of the acceleration horizon. They are analogous to the de Sitter QNMs found in \cite{Jansen:2017oag,Cardoso:2017soq,Destounis:2019hca}. Our numerics suggest that the dominant mode of this family is obtained when $m_0=0$ \cite{Destounis:2020pjk}.

The final family of modes is the near extremal (NE) family (in green in Fig. \ref{QNM} and \ref{beta}) which consists of purely imaginary modes and dominates the ringdown when the event and CH approach each other. This family asymptotes the NE modes of RN \cite{Hod:2017gvn} as $\alpha\rightarrow 0$ and vanishes at extremality. Similar modes have been found in RNdS for scalar \cite{Cardoso:2017soq,Cardoso:2018nvb,Dias:2018ufh} and fermionic perturbations \cite{Destounis:2018qnb}. In our case, the dominant mode  is obtained when $m_0=0$ \cite{Destounis:2020pjk}.

For completeness, in the right panel of Fig. \ref{QNM}, we provide the temporal response of scalar perturbations for three distinct cases, chosen so that a different family dominates in each case. We observe that the late-time behavior of scalar fields on the charged $C-$metric is, indeed, exponential, while the extracted QNMs at late times match accurately the dominant QNMs derived from the spectral analysis. The parameters for these examples are only indicative, as we find that \eqref{expo} holds for various cases throughout the sub-extremal parameter space \cite{Destounis:2020pjk}.

In Fig. \ref{beta}, we present the ratio $-\text{Im}(\omega)/\kappa_-$ determined by the dominant QNMs of each family, in near-extremally-charged accelerating BHs, for various acceleration parameters $\alpha M$. Then, $\beta$ is obtained from the smallest contribution of all families of modes (cf. eq. \eqref{def-beta}). For slowly accelerating charged BHs, the $m_0=0$ acceleration modes dominate throughout most of the sub-extremal parameter space. On the other hand, for sufficiently large acceleration the high-frequency PS modes dominate. Independently of the acceleration parameter though, there is always a small, but finite, region in the parameter space, when $Q\rightarrow M$, where the $m_0=0$ NE family of modes dominates. 

For all cases shown, $\beta$ exceeds $1/2$. In fact, $\beta$ would diverge at extremality if only the PS and acceleration modes were present. However, the NE modes take over to keep $\beta$ from exceeding unity. Nevertheless, the existence of regions in the parameter space of near-extremally-charged accelerating BHs where $\beta>1/2$, indicates a potential violation of SCC.

\section{Conclusions} Until now, BHs without a cosmological constant are known to have unstable Cauchy horizons and satisfy SCC.
In those cases, linear perturbations decay slowly enough \cite{Price:1971fb} to guarantee that the exponential blueshift effect, triggered at the CH, will turn this region into a mass-inflation singularity, where the field equations cease to make sense \cite{Poisson_PRL,Poisson:1990eh}. 

A recent study of linear scalar perturbations in the charged $C-$metric \cite{Destounis:2020pjk} revealed that the existence of acceleration horizons leads to the exponential decay of perturbations which, in turn, are controlled by the dominant QNM at late times, even though the spacetime does not possess a cosmological constant. This can be understood from the fact that the asymptotic region is causally disconnected from the region considered for the evolution of the wave equation and the initial data considered resemble those used for the wave equation in RNdS spacetimes.

In this article, we exploit this late-time behavior of perturbations to test the modern formulation of SCC \cite{Christodoulou:2008nj}, using QNMs \cite{Cardoso:2017soq}, in accelerating, charged BHs.

We have provided robust numerical evidence supporting the fact that the CH of near-extremally-charged accelerating BHs is stable against linear scalar field perturbations, and thus, SCC may be violated. This result is in agreement with the prediction that charged accelerating BHs may violate SCC near extremality \cite{Horowitz:1996yb} and disproves the conjecture of \cite{Brady:1998au}.

Although accelerating BHs without a cosmological constant may violate SCC, the charged $C-$metric describes an idealized scenario where BHs are uniformly accelerating away from each other indefinitely. While such accelerations may approximately occur in particle accelerators, a long lasting acceleration horizon does not seem to be present in BHs residing in binary mergers. The absence of an acceleration horizon would, then, lead to the usual power-law decay of scalar perturbations in asymptotically flat BHs, and SCC would be restored. In any case, the formation  and evolution of cosmic strings \cite{Gibbons:1990gp}, predicted by quantum field theory and string theory, can provide the perpetual energy needed for a BH to form an acceleration horizon.

Finally, if one considers that SCC has been an important mathematical conjecture and a way to test classical GR and its limits, then its potential violation in the charged $C-$metric, which is an exact solution of GR generalizing the RN solution, deserves further investigation.


\noindent{\bf{\em Acknowledgements.}} We thank Jo\~ao Costa and Vitor Cardoso for useful comments. KD acknowledges financial support and hospitality from CAMGSD, IST, Univ. Lisboa where this project was initiated. FCM thanks support from CAMGSD, IST, Univ. Lisboa, and CMAT, Univ. Minho, through FCT funds UID/MAT/04459/2019 and Est-OE/MAT/UIDB/00013/2020, respectively, and FEDER Funds COMPETE.

\bibliography{SCC,references}

\begin{thebibliography}{98}%
\makeatletter
\providecommand \@ifxundefined [1]{%
 \@ifx{#1\undefined}
}%
\providecommand \@ifnum [1]{%
 \ifnum #1\expandafter \@firstoftwo
 \else \expandafter \@secondoftwo
 \fi
}%
\providecommand \@ifx [1]{%
 \ifx #1\expandafter \@firstoftwo
 \else \expandafter \@secondoftwo
 \fi
}%
\providecommand \natexlab [1]{#1}%
\providecommand \enquote  [1]{``#1''}%
\providecommand \bibnamefont  [1]{#1}%
\providecommand \bibfnamefont [1]{#1}%
\providecommand \citenamefont [1]{#1}%
\providecommand \href@noop [0]{\@secondoftwo}%
\providecommand \href [0]{\begingroup \@sanitize@url \@href}%
\providecommand \@href[1]{\@@startlink{#1}\@@href}%
\providecommand \@@href[1]{\endgroup#1\@@endlink}%
\providecommand \@sanitize@url [0]{\catcode `\\12\catcode `\$12\catcode
  `\&12\catcode `\#12\catcode `\^12\catcode `\_12\catcode `\%12\relax}%
\providecommand \@@startlink[1]{}%
\providecommand \@@endlink[0]{}%
\providecommand \url  [0]{\begingroup\@sanitize@url \@url }%
\providecommand \@url [1]{\endgroup\@href {#1}{\urlprefix }}%
\providecommand \urlprefix  [0]{URL }%
\providecommand \Eprint [0]{\href }%
\providecommand \doibase [0]{http://dx.doi.org/}%
\providecommand \selectlanguage [0]{\@gobble}%
\providecommand \bibinfo  [0]{\@secondoftwo}%
\providecommand \bibfield  [0]{\@secondoftwo}%
\providecommand \translation [1]{[#1]}%
\providecommand \BibitemOpen [0]{}%
\providecommand \bibitemStop [0]{}%
\providecommand \bibitemNoStop [0]{.\EOS\space}%
\providecommand \EOS [0]{\spacefactor3000\relax}%
\providecommand \BibitemShut  [1]{\csname bibitem#1\endcsname}%
\let\auto@bib@innerbib\@empty
\bibitem [{\citenamefont {Dafermos}(2005)}]{Dafermos:2003wr}%
  \BibitemOpen
  \bibfield  {author} {\bibinfo {author} {\bibfnamefont {M.}~\bibnamefont
  {Dafermos}},\ }\href@noop {} {\bibfield  {journal} {\bibinfo  {journal}
  {Commun. Pure Appl. Math.}\ }\textbf {\bibinfo {volume} {58}},\ \bibinfo
  {pages} {0445} (\bibinfo {year} {2005})},\ \Eprint
  {http://arxiv.org/abs/gr-qc/0307013} {arXiv:gr-qc/0307013 [gr-qc]}
  \BibitemShut {NoStop}%
\bibitem [{\citenamefont {Ori}(2000)}]{Ori:2000fi}%
  \BibitemOpen
  \bibfield  {author} {\bibinfo {author} {\bibfnamefont {A.}~\bibnamefont
  {Ori}},\ }\href {\doibase 10.1103/PhysRevD.61.064016} {\bibfield  {journal}
  {\bibinfo  {journal} {Phys. Rev.}\ }\textbf {\bibinfo {volume} {D61}},\
  \bibinfo {pages} {064016} (\bibinfo {year} {2000})}\BibitemShut {NoStop}%
\bibitem [{\citenamefont {Sbierski}(2018)}]{Sbierski:2015nta}%
  \BibitemOpen
  \bibfield  {author} {\bibinfo {author} {\bibfnamefont {J.}~\bibnamefont
  {Sbierski}},\ }\href {\doibase 10.4310/jdg/1518490820} {\bibfield  {journal}
  {\bibinfo  {journal} {J. Diff. Geom.}\ }\textbf {\bibinfo {volume} {108}},\
  \bibinfo {pages} {319} (\bibinfo {year} {2018})},\ \Eprint
  {http://arxiv.org/abs/1507.00601} {arXiv:1507.00601 [gr-qc]} \BibitemShut
  {NoStop}%
\bibitem [{\citenamefont {Hawking}\ and\ \citenamefont
  {Ellis}(2011)}]{Hawking:1973uf}%
  \BibitemOpen
  \bibfield  {author} {\bibinfo {author} {\bibfnamefont {S.~W.}\ \bibnamefont
  {Hawking}}\ and\ \bibinfo {author} {\bibfnamefont {G.~F.~R.}\ \bibnamefont
  {Ellis}},\ }\href {\doibase 10.1017/CBO9780511524646} {\emph {\bibinfo
  {title} {{The Large Scale Structure of Space-Time}}}},\ Cambridge Monographs
  on Mathematical Physics\ (\bibinfo  {publisher} {Cambridge University
  Press},\ \bibinfo {year} {2011})\BibitemShut {NoStop}%
\bibitem [{\citenamefont {Kokkotas}\ and\ \citenamefont
  {Schmidt}(1999)}]{Kokkotas:1999bd}%
  \BibitemOpen
  \bibfield  {author} {\bibinfo {author} {\bibfnamefont {K.~D.}\ \bibnamefont
  {Kokkotas}}\ and\ \bibinfo {author} {\bibfnamefont {B.~G.}\ \bibnamefont
  {Schmidt}},\ }\href {\doibase 10.12942/lrr-1999-2} {\bibfield  {journal}
  {\bibinfo  {journal} {Living Rev. Rel.}\ }\textbf {\bibinfo {volume} {2}},\
  \bibinfo {pages} {2} (\bibinfo {year} {1999})},\ \Eprint
  {http://arxiv.org/abs/gr-qc/9909058} {arXiv:gr-qc/9909058 [gr-qc]}
  \BibitemShut {NoStop}%
\bibitem [{\citenamefont {Berti}\ \emph {et~al.}(2009)\citenamefont {Berti},
  \citenamefont {Cardoso},\ and\ \citenamefont {Starinets}}]{Berti:2009kk}%
  \BibitemOpen
  \bibfield  {author} {\bibinfo {author} {\bibfnamefont {E.}~\bibnamefont
  {Berti}}, \bibinfo {author} {\bibfnamefont {V.}~\bibnamefont {Cardoso}}, \
  and\ \bibinfo {author} {\bibfnamefont {A.~O.}\ \bibnamefont {Starinets}},\
  }\href {\doibase 10.1088/0264-9381/26/16/163001} {\bibfield  {journal}
  {\bibinfo  {journal} {Class. Quant. Grav.}\ }\textbf {\bibinfo {volume}
  {26}},\ \bibinfo {pages} {163001} (\bibinfo {year} {2009})},\ \Eprint
  {http://arxiv.org/abs/0905.2975} {arXiv:0905.2975 [gr-qc]} \BibitemShut
  {NoStop}%
\bibitem [{\citenamefont {Konoplya}\ and\ \citenamefont
  {Zhidenko}(2011)}]{Konoplya:2011qq}%
  \BibitemOpen
  \bibfield  {author} {\bibinfo {author} {\bibfnamefont {R.~A.}\ \bibnamefont
  {Konoplya}}\ and\ \bibinfo {author} {\bibfnamefont {A.}~\bibnamefont
  {Zhidenko}},\ }\href {\doibase 10.1103/RevModPhys.83.793} {\bibfield
  {journal} {\bibinfo  {journal} {Rev. Mod. Phys.}\ }\textbf {\bibinfo {volume}
  {83}},\ \bibinfo {pages} {793} (\bibinfo {year} {2011})},\ \Eprint
  {http://arxiv.org/abs/1102.4014} {arXiv:1102.4014 [gr-qc]} \BibitemShut
  {NoStop}%
\bibitem [{\citenamefont {Simpson}\ and\ \citenamefont
  {Penrose}(1973)}]{Simpson:1973ua}%
  \BibitemOpen
  \bibfield  {author} {\bibinfo {author} {\bibfnamefont {M.}~\bibnamefont
  {Simpson}}\ and\ \bibinfo {author} {\bibfnamefont {R.}~\bibnamefont
  {Penrose}},\ }\href {\doibase 10.1007/BF00792069} {\bibfield  {journal}
  {\bibinfo  {journal} {Int. J. Theor. Phys.}\ }\textbf {\bibinfo {volume}
  {7}},\ \bibinfo {pages} {183} (\bibinfo {year} {1973})}\BibitemShut {NoStop}%
\bibitem [{\citenamefont {Chandrasekhar}\ and\ \citenamefont
  {Hartle}(1982)}]{Hartle}%
  \BibitemOpen
  \bibfield  {author} {\bibinfo {author} {\bibfnamefont {S.}~\bibnamefont
  {Chandrasekhar}}\ and\ \bibinfo {author} {\bibfnamefont {J.~B.}\ \bibnamefont
  {Hartle}},\ }\href {http://www.jstor.org/stable/2397225} {\bibfield
  {journal} {\bibinfo  {journal} {Proceedings of the Royal Society of London.
  Series A, Mathematical and Physical Sciences}\ }\textbf {\bibinfo {volume}
  {384}},\ \bibinfo {pages} {301} (\bibinfo {year} {1982})}\BibitemShut
  {NoStop}%
\bibitem [{\citenamefont {Poisson}\ and\ \citenamefont
  {Israel}(1989)}]{Poisson_PRL}%
  \BibitemOpen
  \bibfield  {author} {\bibinfo {author} {\bibfnamefont {E.}~\bibnamefont
  {Poisson}}\ and\ \bibinfo {author} {\bibfnamefont {W.}~\bibnamefont
  {Israel}},\ }\href {\doibase 10.1103/PhysRevLett.63.1663} {\bibfield
  {journal} {\bibinfo  {journal} {Phys. Rev. Lett.}\ }\textbf {\bibinfo
  {volume} {63}},\ \bibinfo {pages} {1663} (\bibinfo {year}
  {1989})}\BibitemShut {NoStop}%
\bibitem [{\citenamefont {Poisson}\ and\ \citenamefont
  {Israel}(1990)}]{Poisson:1990eh}%
  \BibitemOpen
  \bibfield  {author} {\bibinfo {author} {\bibfnamefont {E.}~\bibnamefont
  {Poisson}}\ and\ \bibinfo {author} {\bibfnamefont {W.}~\bibnamefont
  {Israel}},\ }\href {\doibase 10.1103/PhysRevD.41.1796} {\bibfield  {journal}
  {\bibinfo  {journal} {Phys. Rev.}\ }\textbf {\bibinfo {volume} {D41}},\
  \bibinfo {pages} {1796} (\bibinfo {year} {1990})}\BibitemShut {NoStop}%
\bibitem [{\citenamefont {Price}(1972)}]{Price:1971fb}%
  \BibitemOpen
  \bibfield  {author} {\bibinfo {author} {\bibfnamefont {R.~H.}\ \bibnamefont
  {Price}},\ }\href {\doibase 10.1103/PhysRevD.5.2419} {\bibfield  {journal}
  {\bibinfo  {journal} {Phys. Rev.}\ }\textbf {\bibinfo {volume} {D5}},\
  \bibinfo {pages} {2419} (\bibinfo {year} {1972})}\BibitemShut {NoStop}%
\bibitem [{\citenamefont {Leaver}(1985)}]{Leaver:1985ax}%
  \BibitemOpen
  \bibfield  {author} {\bibinfo {author} {\bibfnamefont {E.~W.}\ \bibnamefont
  {Leaver}},\ }\href {\doibase 10.1098/rspa.1985.0119} {\bibfield  {journal}
  {\bibinfo  {journal} {Proc. Roy. Soc. Lond.}\ }\textbf {\bibinfo {volume}
  {A402}},\ \bibinfo {pages} {285} (\bibinfo {year} {1985})}\BibitemShut
  {NoStop}%
\bibitem [{\citenamefont {Dafermos}\ and\ \citenamefont
  {Rodnianski}(2005)}]{Dafermos:2003yw}%
  \BibitemOpen
  \bibfield  {author} {\bibinfo {author} {\bibfnamefont {M.}~\bibnamefont
  {Dafermos}}\ and\ \bibinfo {author} {\bibfnamefont {I.}~\bibnamefont
  {Rodnianski}},\ }\href {\doibase 10.1007/s00222-005-0450-3} {\bibfield
  {journal} {\bibinfo  {journal} {Invent. Math.}\ }\textbf {\bibinfo {volume}
  {162}},\ \bibinfo {pages} {381} (\bibinfo {year} {2005})},\ \Eprint
  {http://arxiv.org/abs/gr-qc/0309115} {arXiv:gr-qc/0309115 [gr-qc]}
  \BibitemShut {NoStop}%
\bibitem [{\citenamefont {Dafermos}\ and\ \citenamefont
  {Rodnianski}(2010)}]{Dafermos:2010hb}%
  \BibitemOpen
  \bibfield  {author} {\bibinfo {author} {\bibfnamefont {M.}~\bibnamefont
  {Dafermos}}\ and\ \bibinfo {author} {\bibfnamefont {I.}~\bibnamefont
  {Rodnianski}},\ }\href@noop {} {\  (\bibinfo {year} {2010})},\ \Eprint
  {http://arxiv.org/abs/1010.5132} {arXiv:1010.5132 [gr-qc]} \BibitemShut
  {NoStop}%
\bibitem [{\citenamefont {Dafermos}\ \emph {et~al.}(2016)\citenamefont
  {Dafermos}, \citenamefont {Rodnianski},\ and\ \citenamefont
  {Shlapentokh-Rothman}}]{Dafermos:2014cua}%
  \BibitemOpen
  \bibfield  {author} {\bibinfo {author} {\bibfnamefont {M.}~\bibnamefont
  {Dafermos}}, \bibinfo {author} {\bibfnamefont {I.}~\bibnamefont
  {Rodnianski}}, \ and\ \bibinfo {author} {\bibfnamefont {Y.}~\bibnamefont
  {Shlapentokh-Rothman}},\ }\href {\doibase 10.4007/annals.2016.183.3.2}
  {\bibfield  {journal} {\bibinfo  {journal} {Ann. Math.}\ }\textbf {\bibinfo
  {volume} {183}},\ \bibinfo {pages} {787} (\bibinfo {year}
  {2016})}\BibitemShut {NoStop}%
\bibitem [{\citenamefont {Gundlach}\ \emph
  {et~al.}(1994{\natexlab{a}})\citenamefont {Gundlach}, \citenamefont {Price},\
  and\ \citenamefont {Pullin}}]{Gundlach1}%
  \BibitemOpen
  \bibfield  {author} {\bibinfo {author} {\bibfnamefont {C.}~\bibnamefont
  {Gundlach}}, \bibinfo {author} {\bibfnamefont {R.~H.}\ \bibnamefont {Price}},
  \ and\ \bibinfo {author} {\bibfnamefont {J.}~\bibnamefont {Pullin}},\ }\href
  {\doibase 10.1103/PhysRevD.49.883} {\bibfield  {journal} {\bibinfo  {journal}
  {Phys. Rev. D}\ }\textbf {\bibinfo {volume} {49}},\ \bibinfo {pages} {883}
  (\bibinfo {year} {1994}{\natexlab{a}})}\BibitemShut {NoStop}%
\bibitem [{\citenamefont {Gundlach}\ \emph
  {et~al.}(1994{\natexlab{b}})\citenamefont {Gundlach}, \citenamefont {Price},\
  and\ \citenamefont {Pullin}}]{Gundlach2}%
  \BibitemOpen
  \bibfield  {author} {\bibinfo {author} {\bibfnamefont {C.}~\bibnamefont
  {Gundlach}}, \bibinfo {author} {\bibfnamefont {R.~H.}\ \bibnamefont {Price}},
  \ and\ \bibinfo {author} {\bibfnamefont {J.}~\bibnamefont {Pullin}},\ }\href
  {\doibase 10.1103/PhysRevD.49.890} {\bibfield  {journal} {\bibinfo  {journal}
  {Phys. Rev. D}\ }\textbf {\bibinfo {volume} {49}},\ \bibinfo {pages} {890}
  (\bibinfo {year} {1994}{\natexlab{b}})}\BibitemShut {NoStop}%
\bibitem [{\citenamefont {{Penrose}}(1969)}]{Penrose69}%
  \BibitemOpen
  \bibfield  {author} {\bibinfo {author} {\bibfnamefont {R.}~\bibnamefont
  {{Penrose}}},\ }\href@noop {} {\bibfield  {journal} {\bibinfo  {journal}
  {Nuovo Cimento Rivista Serie}\ }\textbf {\bibinfo {volume} {1}} (\bibinfo
  {year} {1969})}\BibitemShut {NoStop}%
\bibitem [{\citenamefont {Christodoulou}(1999)}]{christodoulou1999instability}%
  \BibitemOpen
  \bibfield  {author} {\bibinfo {author} {\bibfnamefont {D.}~\bibnamefont
  {Christodoulou}},\ }\href {\doibase 10.2307/121023} {\bibfield  {journal}
  {\bibinfo  {journal} {Ann. of Math.}\ }\textbf {\bibinfo {volume} {144}},\
  \bibinfo {pages} {183} (\bibinfo {year} {1999})}\BibitemShut {NoStop}%
\bibitem [{\citenamefont {Dafermos}(2014)}]{Dafermos:2012np}%
  \BibitemOpen
  \bibfield  {author} {\bibinfo {author} {\bibfnamefont {M.}~\bibnamefont
  {Dafermos}},\ }\href {\doibase 10.1007/s00220-014-2063-4} {\bibfield
  {journal} {\bibinfo  {journal} {Commun. Math. Phys.}\ }\textbf {\bibinfo
  {volume} {332}},\ \bibinfo {pages} {729} (\bibinfo {year} {2014})},\ \Eprint
  {http://arxiv.org/abs/1201.1797} {arXiv:1201.1797 [gr-qc]} \BibitemShut
  {NoStop}%
\bibitem [{\citenamefont {Hintz}\ and\ \citenamefont
  {Vasy}(2017)}]{Hintz:2015jkj}%
  \BibitemOpen
  \bibfield  {author} {\bibinfo {author} {\bibfnamefont {P.}~\bibnamefont
  {Hintz}}\ and\ \bibinfo {author} {\bibfnamefont {A.}~\bibnamefont {Vasy}},\
  }\href {\doibase 10.1063/1.4996575} {\bibfield  {journal} {\bibinfo
  {journal} {J. Math. Phys.}\ }\textbf {\bibinfo {volume} {58}},\ \bibinfo
  {pages} {081509} (\bibinfo {year} {2017})},\ \Eprint
  {http://arxiv.org/abs/1512.08004} {arXiv:1512.08004 [math.AP]} \BibitemShut
  {NoStop}%
\bibitem [{\citenamefont {Hintz}\ and\ \citenamefont
  {Vasy}(2018)}]{Hintz:2016gwb}%
  \BibitemOpen
  \bibfield  {author} {\bibinfo {author} {\bibfnamefont {P.}~\bibnamefont
  {Hintz}}\ and\ \bibinfo {author} {\bibfnamefont {A.}~\bibnamefont {Vasy}},\
  }\href@noop {} {\bibfield  {journal} {\bibinfo  {journal} {Acta Mathematica}\
  }\textbf {\bibinfo {volume} {220}},\ \bibinfo {pages} {1} (\bibinfo {year}
  {2018})}\BibitemShut {NoStop}%
\bibitem [{\citenamefont {Hintz}(2018)}]{Hintz:2016jak}%
  \BibitemOpen
  \bibfield  {author} {\bibinfo {author} {\bibfnamefont {P.}~\bibnamefont
  {Hintz}},\ }\href {\doibase 10.1007/s40818-018-0047-y} {\bibfield  {journal}
  {\bibinfo  {journal} {Annals of PDE}\ } (\bibinfo {year} {2018}),\
  10.1007/s40818-018-0047-y},\ \Eprint {http://arxiv.org/abs/1612.04489}
  {arXiv:1612.04489 [math.AP]} \BibitemShut {NoStop}%
\bibitem [{\citenamefont {Bony}\ and\ \citenamefont
  {H{\"a}fner}(2008)}]{BonyHaefner}%
  \BibitemOpen
  \bibfield  {author} {\bibinfo {author} {\bibfnamefont {J.-F.}\ \bibnamefont
  {Bony}}\ and\ \bibinfo {author} {\bibfnamefont {D.}~\bibnamefont
  {H{\"a}fner}},\ }\href@noop {} {\bibfield  {journal} {\bibinfo  {journal}
  {Communications in Mathematical Physics}\ }\textbf {\bibinfo {volume}
  {282}},\ \bibinfo {pages} {697} (\bibinfo {year} {2008})}\BibitemShut
  {NoStop}%
\bibitem [{\citenamefont {Barreto}\ and\ \citenamefont
  {Zworski}(1997)}]{saBarreto}%
  \BibitemOpen
  \bibfield  {author} {\bibinfo {author} {\bibfnamefont {S.}~\bibnamefont
  {Barreto}}\ and\ \bibinfo {author} {\bibfnamefont {M.}~\bibnamefont
  {Zworski}},\ }\href@noop {} {\bibfield  {journal} {\bibinfo  {journal}
  {Math.\ Res.\ Lett.}\ }\textbf {\bibinfo {volume} {4}},\ \bibinfo {pages}
  {103} (\bibinfo {year} {1997})}\BibitemShut {NoStop}%
\bibitem [{\citenamefont {Dyatlov}(2012)}]{Dyatlov:2011jd}%
  \BibitemOpen
  \bibfield  {author} {\bibinfo {author} {\bibfnamefont {S.}~\bibnamefont
  {Dyatlov}},\ }\href {\doibase 10.1007/s00023-012-0159-y} {\bibfield
  {journal} {\bibinfo  {journal} {Annales Henri Poincare}\ }\textbf {\bibinfo
  {volume} {13}},\ \bibinfo {pages} {1101} (\bibinfo {year} {2012})},\ \Eprint
  {http://arxiv.org/abs/1101.1260} {arXiv:1101.1260 [math.AP]} \BibitemShut
  {NoStop}%
\bibitem [{\citenamefont {Dyatlov}(2015)}]{Dyatlov:2013hba}%
  \BibitemOpen
  \bibfield  {author} {\bibinfo {author} {\bibfnamefont {S.}~\bibnamefont
  {Dyatlov}},\ }\href {\doibase 10.1007/s00220-014-2255-y} {\bibfield
  {journal} {\bibinfo  {journal} {Commun. Math. Phys.}\ }\textbf {\bibinfo
  {volume} {335}},\ \bibinfo {pages} {1445} (\bibinfo {year} {2015})},\ \Eprint
  {http://arxiv.org/abs/1305.1723} {arXiv:1305.1723 [gr-qc]} \BibitemShut
  {NoStop}%
\bibitem [{\citenamefont {Costa}\ \emph
  {et~al.}(2015{\natexlab{a}})\citenamefont {Costa}, \citenamefont {Gir\~{a}o},
  \citenamefont {Nat\'{a}rio},\ and\ \citenamefont {Silva}}]{Costa:2014yha}%
  \BibitemOpen
  \bibfield  {author} {\bibinfo {author} {\bibfnamefont {J.~L.}\ \bibnamefont
  {Costa}}, \bibinfo {author} {\bibfnamefont {P.~M.}\ \bibnamefont
  {Gir\~{a}o}}, \bibinfo {author} {\bibfnamefont {J.}~\bibnamefont
  {Nat\'{a}rio}}, \ and\ \bibinfo {author} {\bibfnamefont {J.~D.}\ \bibnamefont
  {Silva}},\ }\href {\doibase 10.1088/0264-9381/32/1/015017} {\bibfield
  {journal} {\bibinfo  {journal} {Class. Quant. Grav.}\ }\textbf {\bibinfo
  {volume} {32}},\ \bibinfo {pages} {015017} (\bibinfo {year}
  {2015}{\natexlab{a}})},\ \Eprint {http://arxiv.org/abs/1406.7245}
  {arXiv:1406.7245 [gr-qc]} \BibitemShut {NoStop}%
\bibitem [{\citenamefont {Costa}\ \emph
  {et~al.}(2015{\natexlab{b}})\citenamefont {Costa}, \citenamefont {Gir\~{a}o},
  \citenamefont {Nat\'{a}rio},\ and\ \citenamefont {Silva}}]{Costa:2014zha}%
  \BibitemOpen
  \bibfield  {author} {\bibinfo {author} {\bibfnamefont {J.~L.}\ \bibnamefont
  {Costa}}, \bibinfo {author} {\bibfnamefont {P.~M.}\ \bibnamefont
  {Gir\~{a}o}}, \bibinfo {author} {\bibfnamefont {J.}~\bibnamefont
  {Nat\'{a}rio}}, \ and\ \bibinfo {author} {\bibfnamefont {J.~D.}\ \bibnamefont
  {Silva}},\ }\href {\doibase 10.1007/s00220-015-2433-6} {\bibfield  {journal}
  {\bibinfo  {journal} {Commun. Math. Phys.}\ }\textbf {\bibinfo {volume}
  {339}},\ \bibinfo {pages} {903} (\bibinfo {year} {2015}{\natexlab{b}})},\
  \Eprint {http://arxiv.org/abs/1406.7253} {arXiv:1406.7253 [gr-qc]}
  \BibitemShut {NoStop}%
\bibitem [{\citenamefont {Costa}\ \emph {et~al.}(2017)\citenamefont {Costa},
  \citenamefont {Gir\~{a}o}, \citenamefont {Nat\'{a}rio},\ and\ \citenamefont
  {Silva}}]{Costa:2014aia}%
  \BibitemOpen
  \bibfield  {author} {\bibinfo {author} {\bibfnamefont {J.~L.}\ \bibnamefont
  {Costa}}, \bibinfo {author} {\bibfnamefont {P.~M.}\ \bibnamefont
  {Gir\~{a}o}}, \bibinfo {author} {\bibfnamefont {J.}~\bibnamefont
  {Nat\'{a}rio}}, \ and\ \bibinfo {author} {\bibfnamefont {J.~D.}\ \bibnamefont
  {Silva}},\ }\href {\doibase 10.1103/PhysRevD.69.064033} {\bibfield  {journal}
  {\bibinfo  {journal} {Ann. PDE}\ }\textbf {\bibinfo {volume} {3: 8}},\
  \bibinfo {pages} {064033} (\bibinfo {year} {2017})}\BibitemShut {NoStop}%
\bibitem [{\citenamefont {Costa}\ \emph {et~al.}(2018)\citenamefont {Costa},
  \citenamefont {Girão}, \citenamefont {Natário},\ and\ \citenamefont
  {Silva}}]{Costa:2017tjc}%
  \BibitemOpen
  \bibfield  {author} {\bibinfo {author} {\bibfnamefont {J.~L.}\ \bibnamefont
  {Costa}}, \bibinfo {author} {\bibfnamefont {P.~M.}\ \bibnamefont {Girão}},
  \bibinfo {author} {\bibfnamefont {J.}~\bibnamefont {Natário}}, \ and\
  \bibinfo {author} {\bibfnamefont {J.~D.}\ \bibnamefont {Silva}},\ }\href
  {\doibase 10.1007/s00220-018-3122-z} {\bibfield  {journal} {\bibinfo
  {journal} {Commun. Math. Phys.}\ }\textbf {\bibinfo {volume} {361}},\
  \bibinfo {pages} {289} (\bibinfo {year} {2018})},\ \Eprint
  {http://arxiv.org/abs/1707.08975} {arXiv:1707.08975 [gr-qc]} \BibitemShut
  {NoStop}%
\bibitem [{\citenamefont {Ori}(1991)}]{Ori:1991zz}%
  \BibitemOpen
  \bibfield  {author} {\bibinfo {author} {\bibfnamefont {A.}~\bibnamefont
  {Ori}},\ }\href {\doibase 10.1103/PhysRevLett.67.789} {\bibfield  {journal}
  {\bibinfo  {journal} {Phys. Rev. Lett.}\ }\textbf {\bibinfo {volume} {67}},\
  \bibinfo {pages} {789} (\bibinfo {year} {1991})}\BibitemShut {NoStop}%
\bibitem [{\citenamefont {Dafermos}\ and\ \citenamefont
  {Luk}(2017)}]{Dafermos:2017dbw}%
  \BibitemOpen
  \bibfield  {author} {\bibinfo {author} {\bibfnamefont {M.}~\bibnamefont
  {Dafermos}}\ and\ \bibinfo {author} {\bibfnamefont {J.}~\bibnamefont {Luk}},\
  }\href@noop {} {\  (\bibinfo {year} {2017})},\ \Eprint
  {http://arxiv.org/abs/1710.01722} {arXiv:1710.01722 [gr-qc]} \BibitemShut
  {NoStop}%
\bibitem [{\citenamefont {Cardoso}\ \emph
  {et~al.}(2018{\natexlab{a}})\citenamefont {Cardoso}, \citenamefont {Costa},
  \citenamefont {Destounis}, \citenamefont {Hintz},\ and\ \citenamefont
  {Jansen}}]{Cardoso:2017soq}%
  \BibitemOpen
  \bibfield  {author} {\bibinfo {author} {\bibfnamefont {V.}~\bibnamefont
  {Cardoso}}, \bibinfo {author} {\bibfnamefont {J.~L.}\ \bibnamefont {Costa}},
  \bibinfo {author} {\bibfnamefont {K.}~\bibnamefont {Destounis}}, \bibinfo
  {author} {\bibfnamefont {P.}~\bibnamefont {Hintz}}, \ and\ \bibinfo {author}
  {\bibfnamefont {A.}~\bibnamefont {Jansen}},\ }\href {\doibase
  10.1103/PhysRevLett.120.031103} {\bibfield  {journal} {\bibinfo  {journal}
  {Phys. Rev. Lett.}\ }\textbf {\bibinfo {volume} {120}},\ \bibinfo {pages}
  {031103} (\bibinfo {year} {2018}{\natexlab{a}})},\ \Eprint
  {http://arxiv.org/abs/1711.10502} {arXiv:1711.10502 [gr-qc]} \BibitemShut
  {NoStop}%
\bibitem [{\citenamefont {Costa}\ and\ \citenamefont
  {Franzen}(2017)}]{CostaFranzen}%
  \BibitemOpen
  \bibfield  {author} {\bibinfo {author} {\bibfnamefont {J.~L.}\ \bibnamefont
  {Costa}}\ and\ \bibinfo {author} {\bibfnamefont {A.~T.}\ \bibnamefont
  {Franzen}},\ }\href {\doibase 10.1007/s00023-017-0592-z} {\bibfield
  {journal} {\bibinfo  {journal} {Ann. Henri Poincar\'{e}}\ }\textbf {\bibinfo
  {volume} {18}},\ \bibinfo {pages} {3371} (\bibinfo {year} {2017})},\ \Eprint
  {http://arxiv.org/abs/1607.01018} {arXiv:1607.01018 [gr-qc]} \BibitemShut
  {NoStop}%
\bibitem [{\citenamefont {Klainerman}\ \emph {et~al.}(2012)\citenamefont
  {Klainerman}, \citenamefont {Rodnianski},\ and\ \citenamefont
  {Szeftel}}]{Klainerman:2012wt}%
  \BibitemOpen
  \bibfield  {author} {\bibinfo {author} {\bibfnamefont {S.}~\bibnamefont
  {Klainerman}}, \bibinfo {author} {\bibfnamefont {I.}~\bibnamefont
  {Rodnianski}}, \ and\ \bibinfo {author} {\bibfnamefont {J.}~\bibnamefont
  {Szeftel}},\ }\href@noop {} {\bibfield  {journal} {\bibinfo  {journal}
  {Invent. Math.}\ }\textbf {\bibinfo {volume} {202}},\ \bibinfo {pages} {91}
  (\bibinfo {year} {2012})},\ \Eprint {http://arxiv.org/abs/1204.1767}
  {arXiv:1204.1767 [math.AP]} \BibitemShut {NoStop}%
\bibitem [{\citenamefont {Cardoso}\ \emph
  {et~al.}(2018{\natexlab{b}})\citenamefont {Cardoso}, \citenamefont {Costa},
  \citenamefont {Destounis}, \citenamefont {Hintz},\ and\ \citenamefont
  {Jansen}}]{Cardoso:2018nvb}%
  \BibitemOpen
  \bibfield  {author} {\bibinfo {author} {\bibfnamefont {V.}~\bibnamefont
  {Cardoso}}, \bibinfo {author} {\bibfnamefont {J.~L.}\ \bibnamefont {Costa}},
  \bibinfo {author} {\bibfnamefont {K.}~\bibnamefont {Destounis}}, \bibinfo
  {author} {\bibfnamefont {P.}~\bibnamefont {Hintz}}, \ and\ \bibinfo {author}
  {\bibfnamefont {A.}~\bibnamefont {Jansen}},\ }\href {\doibase
  10.1103/PhysRevD.98.104007} {\bibfield  {journal} {\bibinfo  {journal} {Phys.
  Rev.}\ }\textbf {\bibinfo {volume} {D98}},\ \bibinfo {pages} {104007}
  (\bibinfo {year} {2018}{\natexlab{b}})},\ \Eprint
  {http://arxiv.org/abs/1808.03631} {arXiv:1808.03631 [gr-qc]} \BibitemShut
  {NoStop}%
\bibitem [{\citenamefont {Mo}\ \emph {et~al.}(2018)\citenamefont {Mo},
  \citenamefont {Tian}, \citenamefont {Wang}, \citenamefont {Zhang},\ and\
  \citenamefont {Zhong}}]{Zhang1}%
  \BibitemOpen
  \bibfield  {author} {\bibinfo {author} {\bibfnamefont {Y.}~\bibnamefont
  {Mo}}, \bibinfo {author} {\bibfnamefont {Y.}~\bibnamefont {Tian}}, \bibinfo
  {author} {\bibfnamefont {B.}~\bibnamefont {Wang}}, \bibinfo {author}
  {\bibfnamefont {H.}~\bibnamefont {Zhang}}, \ and\ \bibinfo {author}
  {\bibfnamefont {Z.}~\bibnamefont {Zhong}},\ }\href {\doibase
  10.1103/PhysRevD.98.124025} {\bibfield  {journal} {\bibinfo  {journal} {Phys.
  Rev.}\ }\textbf {\bibinfo {volume} {D98}},\ \bibinfo {pages} {124025}
  (\bibinfo {year} {2018})},\ \Eprint {http://arxiv.org/abs/1808.03635}
  {arXiv:1808.03635 [gr-qc]} \BibitemShut {NoStop}%
\bibitem [{\citenamefont {Dias}\ \emph
  {et~al.}(2019{\natexlab{a}})\citenamefont {Dias}, \citenamefont {Reall},\
  and\ \citenamefont {Santos}}]{Dias:2018ufh}%
  \BibitemOpen
  \bibfield  {author} {\bibinfo {author} {\bibfnamefont {O.~J.~C.}\
  \bibnamefont {Dias}}, \bibinfo {author} {\bibfnamefont {H.~S.}\ \bibnamefont
  {Reall}}, \ and\ \bibinfo {author} {\bibfnamefont {J.~E.}\ \bibnamefont
  {Santos}},\ }\href {\doibase 10.1088/1361-6382/aafcf2} {\bibfield  {journal}
  {\bibinfo  {journal} {Class. Quant. Grav.}\ }\textbf {\bibinfo {volume}
  {36}},\ \bibinfo {pages} {045005} (\bibinfo {year} {2019}{\natexlab{a}})},\
  \Eprint {http://arxiv.org/abs/1808.04832} {arXiv:1808.04832 [gr-qc]}
  \BibitemShut {NoStop}%
\bibitem [{\citenamefont {Destounis}(2019{\natexlab{a}})}]{Destounis:2018qnb}%
  \BibitemOpen
  \bibfield  {author} {\bibinfo {author} {\bibfnamefont {K.}~\bibnamefont
  {Destounis}},\ }\href {\doibase 10.1016/j.physletb.2019.06.015} {\bibfield
  {journal} {\bibinfo  {journal} {Phys. Lett.}\ }\textbf {\bibinfo {volume}
  {B795}},\ \bibinfo {pages} {211} (\bibinfo {year} {2019}{\natexlab{a}})},\
  \Eprint {http://arxiv.org/abs/1811.10629} {arXiv:1811.10629 [gr-qc]}
  \BibitemShut {NoStop}%
\bibitem [{\citenamefont {Ge}\ \emph {et~al.}(2019)\citenamefont {Ge},
  \citenamefont {Jiang}, \citenamefont {Wang}, \citenamefont {Zhang},\ and\
  \citenamefont {Zhong}}]{Zhang2}%
  \BibitemOpen
  \bibfield  {author} {\bibinfo {author} {\bibfnamefont {B.}~\bibnamefont
  {Ge}}, \bibinfo {author} {\bibfnamefont {J.}~\bibnamefont {Jiang}}, \bibinfo
  {author} {\bibfnamefont {B.}~\bibnamefont {Wang}}, \bibinfo {author}
  {\bibfnamefont {H.}~\bibnamefont {Zhang}}, \ and\ \bibinfo {author}
  {\bibfnamefont {Z.}~\bibnamefont {Zhong}},\ }\href {\doibase
  10.1007/JHEP01(2019)123} {\bibfield  {journal} {\bibinfo  {journal} {JHEP}\
  }\textbf {\bibinfo {volume} {01}},\ \bibinfo {pages} {123} (\bibinfo {year}
  {2019})},\ \Eprint {http://arxiv.org/abs/1810.12128} {arXiv:1810.12128
  [gr-qc]} \BibitemShut {NoStop}%
\bibitem [{\citenamefont {Dias}\ \emph
  {et~al.}(2018{\natexlab{a}})\citenamefont {Dias}, \citenamefont {Reall},\
  and\ \citenamefont {Santos}}]{Dias:2018etb}%
  \BibitemOpen
  \bibfield  {author} {\bibinfo {author} {\bibfnamefont {O.~J.~C.}\
  \bibnamefont {Dias}}, \bibinfo {author} {\bibfnamefont {H.~S.}\ \bibnamefont
  {Reall}}, \ and\ \bibinfo {author} {\bibfnamefont {J.~E.}\ \bibnamefont
  {Santos}},\ }\href {\doibase 10.1007/JHEP10(2018)001} {\bibfield  {journal}
  {\bibinfo  {journal} {JHEP}\ }\textbf {\bibinfo {volume} {10}},\ \bibinfo
  {pages} {001} (\bibinfo {year} {2018}{\natexlab{a}})},\ \Eprint
  {http://arxiv.org/abs/1808.02895} {arXiv:1808.02895 [gr-qc]} \BibitemShut
  {NoStop}%
\bibitem [{\citenamefont {Dias}\ \emph
  {et~al.}(2018{\natexlab{b}})\citenamefont {Dias}, \citenamefont {Eperon},
  \citenamefont {Reall},\ and\ \citenamefont {Santos}}]{Dias:2018ynt}%
  \BibitemOpen
  \bibfield  {author} {\bibinfo {author} {\bibfnamefont {O.~J.~C.}\
  \bibnamefont {Dias}}, \bibinfo {author} {\bibfnamefont {F.~C.}\ \bibnamefont
  {Eperon}}, \bibinfo {author} {\bibfnamefont {H.~S.}\ \bibnamefont {Reall}}, \
  and\ \bibinfo {author} {\bibfnamefont {J.~E.}\ \bibnamefont {Santos}},\
  }\href {\doibase 10.1103/PhysRevD.97.104060} {\bibfield  {journal} {\bibinfo
  {journal} {Phys. Rev.}\ }\textbf {\bibinfo {volume} {D97}},\ \bibinfo {pages}
  {104060} (\bibinfo {year} {2018}{\natexlab{b}})},\ \Eprint
  {http://arxiv.org/abs/1801.09694} {arXiv:1801.09694 [gr-qc]} \BibitemShut
  {NoStop}%
\bibitem [{\citenamefont {Dafermos}\ and\ \citenamefont
  {Shlapentokh-Rothman}(2018)}]{Dafermos:2018tha}%
  \BibitemOpen
  \bibfield  {author} {\bibinfo {author} {\bibfnamefont {M.}~\bibnamefont
  {Dafermos}}\ and\ \bibinfo {author} {\bibfnamefont {Y.}~\bibnamefont
  {Shlapentokh-Rothman}},\ }\href {\doibase 10.1088/1361-6382/aadbcf}
  {\bibfield  {journal} {\bibinfo  {journal} {Class. Quant. Grav.}\ }\textbf
  {\bibinfo {volume} {35}},\ \bibinfo {pages} {195010} (\bibinfo {year}
  {2018})},\ \Eprint {http://arxiv.org/abs/1805.08764} {arXiv:1805.08764
  [gr-qc]} \BibitemShut {NoStop}%
\bibitem [{\citenamefont {Luna}\ \emph {et~al.}(2019)\citenamefont {Luna},
  \citenamefont {Zilhão}, \citenamefont {Cardoso}, \citenamefont {Costa},\
  and\ \citenamefont {Natário}}]{Luna:2018jfk}%
  \BibitemOpen
  \bibfield  {author} {\bibinfo {author} {\bibfnamefont {R.}~\bibnamefont
  {Luna}}, \bibinfo {author} {\bibfnamefont {M.}~\bibnamefont {Zilhão}},
  \bibinfo {author} {\bibfnamefont {V.}~\bibnamefont {Cardoso}}, \bibinfo
  {author} {\bibfnamefont {J.~L.}\ \bibnamefont {Costa}}, \ and\ \bibinfo
  {author} {\bibfnamefont {J.}~\bibnamefont {Natário}},\ }\href {\doibase
  10.1103/PhysRevD.99.064014} {\bibfield  {journal} {\bibinfo  {journal} {Phys.
  Rev.}\ }\textbf {\bibinfo {volume} {D99}},\ \bibinfo {pages} {064014}
  (\bibinfo {year} {2019})},\ \Eprint {http://arxiv.org/abs/1810.00886}
  {arXiv:1810.00886 [gr-qc]} \BibitemShut {NoStop}%
\bibitem [{\citenamefont {Rahman}\ \emph {et~al.}(2019)\citenamefont {Rahman},
  \citenamefont {Chakraborty}, \citenamefont {SenGupta},\ and\ \citenamefont
  {Sen}}]{Rahman:2018oso}%
  \BibitemOpen
  \bibfield  {author} {\bibinfo {author} {\bibfnamefont {M.}~\bibnamefont
  {Rahman}}, \bibinfo {author} {\bibfnamefont {S.}~\bibnamefont {Chakraborty}},
  \bibinfo {author} {\bibfnamefont {S.}~\bibnamefont {SenGupta}}, \ and\
  \bibinfo {author} {\bibfnamefont {A.~A.}\ \bibnamefont {Sen}},\ }\href
  {\doibase 10.1007/JHEP03(2019)178} {\bibfield  {journal} {\bibinfo  {journal}
  {JHEP}\ }\textbf {\bibinfo {volume} {03}},\ \bibinfo {pages} {178} (\bibinfo
  {year} {2019})},\ \Eprint {http://arxiv.org/abs/1811.08538} {arXiv:1811.08538
  [gr-qc]} \BibitemShut {NoStop}%
\bibitem [{\citenamefont {Liu}\ \emph {et~al.}(2019{\natexlab{a}})\citenamefont
  {Liu}, \citenamefont {Tang}, \citenamefont {Destounis}, \citenamefont {Wang},
  \citenamefont {Papantonopoulos},\ and\ \citenamefont {Zhang}}]{Liu:2019lon}%
  \BibitemOpen
  \bibfield  {author} {\bibinfo {author} {\bibfnamefont {H.}~\bibnamefont
  {Liu}}, \bibinfo {author} {\bibfnamefont {Z.}~\bibnamefont {Tang}}, \bibinfo
  {author} {\bibfnamefont {K.}~\bibnamefont {Destounis}}, \bibinfo {author}
  {\bibfnamefont {B.}~\bibnamefont {Wang}}, \bibinfo {author} {\bibfnamefont
  {E.}~\bibnamefont {Papantonopoulos}}, \ and\ \bibinfo {author} {\bibfnamefont
  {H.}~\bibnamefont {Zhang}},\ }\href {\doibase 10.1007/JHEP03(2019)187}
  {\bibfield  {journal} {\bibinfo  {journal} {JHEP}\ }\textbf {\bibinfo
  {volume} {03}},\ \bibinfo {pages} {187} (\bibinfo {year}
  {2019}{\natexlab{a}})},\ \Eprint {http://arxiv.org/abs/1902.01865}
  {arXiv:1902.01865 [gr-qc]} \BibitemShut {NoStop}%
\bibitem [{\citenamefont {Gwak}(2019)}]{Gwak:2018rba}%
  \BibitemOpen
  \bibfield  {author} {\bibinfo {author} {\bibfnamefont {B.}~\bibnamefont
  {Gwak}},\ }\href {\doibase 10.1140/epjc/s10052-019-7283-5} {\bibfield
  {journal} {\bibinfo  {journal} {Eur. Phys. J. C}\ }\textbf {\bibinfo {volume}
  {79}},\ \bibinfo {pages} {767} (\bibinfo {year} {2019})},\ \Eprint
  {http://arxiv.org/abs/1812.04923} {arXiv:1812.04923 [gr-qc]} \BibitemShut
  {NoStop}%
\bibitem [{\citenamefont {Gim}\ and\ \citenamefont {Gwak}(2019)}]{Gim:2019rkl}%
  \BibitemOpen
  \bibfield  {author} {\bibinfo {author} {\bibfnamefont {Y.}~\bibnamefont
  {Gim}}\ and\ \bibinfo {author} {\bibfnamefont {B.}~\bibnamefont {Gwak}},\
  }\href {\doibase 10.1103/PhysRevD.100.124001} {\bibfield  {journal} {\bibinfo
   {journal} {Phys. Rev. D}\ }\textbf {\bibinfo {volume} {100}},\ \bibinfo
  {pages} {124001} (\bibinfo {year} {2019})}\BibitemShut {NoStop}%
\bibitem [{\citenamefont {Etesi}(2019)}]{Etesi:2019arr}%
  \BibitemOpen
  \bibfield  {author} {\bibinfo {author} {\bibfnamefont {G.}~\bibnamefont
  {Etesi}},\ }\href@noop {} {\  (\bibinfo {year} {2019})},\ \Eprint
  {http://arxiv.org/abs/1905.03952} {arXiv:1905.03952 [gr-qc]} \BibitemShut
  {NoStop}%
\bibitem [{\citenamefont {Rahman}(2019)}]{Rahman:2019uwf}%
  \BibitemOpen
  \bibfield  {author} {\bibinfo {author} {\bibfnamefont {M.}~\bibnamefont
  {Rahman}},\ }\href@noop {} {\  (\bibinfo {year} {2019})},\ \Eprint
  {http://arxiv.org/abs/1905.06675} {arXiv:1905.06675 [gr-qc]} \BibitemShut
  {NoStop}%
\bibitem [{\citenamefont {Guo}\ \emph {et~al.}(2019)\citenamefont {Guo},
  \citenamefont {Liu}, \citenamefont {Kuang},\ and\ \citenamefont
  {Wang}}]{Guo:2019tjy}%
  \BibitemOpen
  \bibfield  {author} {\bibinfo {author} {\bibfnamefont {H.}~\bibnamefont
  {Guo}}, \bibinfo {author} {\bibfnamefont {H.}~\bibnamefont {Liu}}, \bibinfo
  {author} {\bibfnamefont {X.-M.}\ \bibnamefont {Kuang}}, \ and\ \bibinfo
  {author} {\bibfnamefont {B.}~\bibnamefont {Wang}},\ }\href {\doibase
  10.1140/epjc/s10052-019-7416-x} {\bibfield  {journal} {\bibinfo  {journal}
  {Eur. Phys. J. C}\ }\textbf {\bibinfo {volume} {79}},\ \bibinfo {pages} {891}
  (\bibinfo {year} {2019})}\BibitemShut {NoStop}%
\bibitem [{\citenamefont {Dias}\ \emph
  {et~al.}(2019{\natexlab{b}})\citenamefont {Dias}, \citenamefont {Reall},\
  and\ \citenamefont {Santos}}]{Dias:2019ery}%
  \BibitemOpen
  \bibfield  {author} {\bibinfo {author} {\bibfnamefont {O.~J.~C.}\
  \bibnamefont {Dias}}, \bibinfo {author} {\bibfnamefont {H.~S.}\ \bibnamefont
  {Reall}}, \ and\ \bibinfo {author} {\bibfnamefont {J.~E.}\ \bibnamefont
  {Santos}},\ }\href {\doibase 10.1007/JHEP12(2019)097} {\bibfield  {journal}
  {\bibinfo  {journal} {J. High Energ. Phys.}\ }\textbf {\bibinfo {volume}
  {2019}},\ \bibinfo {pages} {97} (\bibinfo {year}
  {2019}{\natexlab{b}})}\BibitemShut {NoStop}%
\bibitem [{\citenamefont {Gan}\ \emph {et~al.}(2019{\natexlab{a}})\citenamefont
  {Gan}, \citenamefont {Guo}, \citenamefont {Wang},\ and\ \citenamefont
  {Wu}}]{Gan:2019jac}%
  \BibitemOpen
  \bibfield  {author} {\bibinfo {author} {\bibfnamefont {Q.}~\bibnamefont
  {Gan}}, \bibinfo {author} {\bibfnamefont {G.}~\bibnamefont {Guo}}, \bibinfo
  {author} {\bibfnamefont {P.}~\bibnamefont {Wang}}, \ and\ \bibinfo {author}
  {\bibfnamefont {H.}~\bibnamefont {Wu}},\ }\href {\doibase
  10.1103/PhysRevD.100.124009} {\bibfield  {journal} {\bibinfo  {journal}
  {Phys. Rev. D}\ }\textbf {\bibinfo {volume} {100}},\ \bibinfo {pages}
  {124009} (\bibinfo {year} {2019}{\natexlab{a}})}\BibitemShut {NoStop}%
\bibitem [{\citenamefont {Destounis}\ \emph {et~al.}(2019)\citenamefont
  {Destounis}, \citenamefont {Fontana}, \citenamefont {Mena},\ and\
  \citenamefont {Papantonopoulos}}]{Destounis:2019omd}%
  \BibitemOpen
  \bibfield  {author} {\bibinfo {author} {\bibfnamefont {K.}~\bibnamefont
  {Destounis}}, \bibinfo {author} {\bibfnamefont {R.~D.}\ \bibnamefont
  {Fontana}}, \bibinfo {author} {\bibfnamefont {F.~C.}\ \bibnamefont {Mena}}, \
  and\ \bibinfo {author} {\bibfnamefont {E.}~\bibnamefont {Papantonopoulos}},\
  }\href {\doibase 10.1007/JHEP10(2019)280} {\bibfield  {journal} {\bibinfo
  {journal} {JHEP}\ }\textbf {\bibinfo {volume} {10}},\ \bibinfo {pages} {280}
  (\bibinfo {year} {2019})},\ \Eprint {http://arxiv.org/abs/1908.09842}
  {arXiv:1908.09842 [gr-qc]} \BibitemShut {NoStop}%
\bibitem [{\citenamefont {Liu}\ \emph {et~al.}(2019{\natexlab{b}})\citenamefont
  {Liu}, \citenamefont {Van~Vooren}, \citenamefont {Zhang},\ and\ \citenamefont
  {Zhong}}]{Liu:2019rbq}%
  \BibitemOpen
  \bibfield  {author} {\bibinfo {author} {\bibfnamefont {X.}~\bibnamefont
  {Liu}}, \bibinfo {author} {\bibfnamefont {S.}~\bibnamefont {Van~Vooren}},
  \bibinfo {author} {\bibfnamefont {H.}~\bibnamefont {Zhang}}, \ and\ \bibinfo
  {author} {\bibfnamefont {Z.}~\bibnamefont {Zhong}},\ }\href {\doibase
  10.1007/JHEP10(2019)186} {\bibfield  {journal} {\bibinfo  {journal} {JHEP}\
  }\textbf {\bibinfo {volume} {10}},\ \bibinfo {pages} {186} (\bibinfo {year}
  {2019}{\natexlab{b}})},\ \Eprint {http://arxiv.org/abs/1909.07904}
  {arXiv:1909.07904 [hep-th]} \BibitemShut {NoStop}%
\bibitem [{\citenamefont {Zhang}\ and\ \citenamefont
  {Zhong}(2019)}]{Zhang:2019nye}%
  \BibitemOpen
  \bibfield  {author} {\bibinfo {author} {\bibfnamefont {H.}~\bibnamefont
  {Zhang}}\ and\ \bibinfo {author} {\bibfnamefont {Z.}~\bibnamefont {Zhong}},\
  }\href@noop {} {\  (\bibinfo {year} {2019})},\ \Eprint
  {http://arxiv.org/abs/1910.01610} {arXiv:1910.01610 [hep-th]} \BibitemShut
  {NoStop}%
\bibitem [{\citenamefont {Chen}\ \emph {et~al.}(2020)\citenamefont {Chen},
  \citenamefont {Gan},\ and\ \citenamefont {Guo}}]{Chen:2019qbz}%
  \BibitemOpen
  \bibfield  {author} {\bibinfo {author} {\bibfnamefont {Y.}~\bibnamefont
  {Chen}}, \bibinfo {author} {\bibfnamefont {Q.}~\bibnamefont {Gan}}, \ and\
  \bibinfo {author} {\bibfnamefont {G.}~\bibnamefont {Guo}},\ }\href {\doibase
  10.1088/1572-9494/ab6912} {\bibfield  {journal} {\bibinfo  {journal} {Commun.
  Theor. Phys.}\ }\textbf {\bibinfo {volume} {72}},\ \bibinfo {pages} {035405}
  (\bibinfo {year} {2020})},\ \Eprint {http://arxiv.org/abs/1911.06628}
  {arXiv:1911.06628 [gr-qc]} \BibitemShut {NoStop}%
\bibitem [{\citenamefont {Mishra}\ and\ \citenamefont
  {Chakraborty}(2020)}]{Mishra:2019ged}%
  \BibitemOpen
  \bibfield  {author} {\bibinfo {author} {\bibfnamefont {A.~K.}\ \bibnamefont
  {Mishra}}\ and\ \bibinfo {author} {\bibfnamefont {S.}~\bibnamefont
  {Chakraborty}},\ }\href {\doibase 10.1103/PhysRevD.101.064041} {\bibfield
  {journal} {\bibinfo  {journal} {Phys. Rev. D}\ }\textbf {\bibinfo {volume}
  {101}},\ \bibinfo {pages} {064041} (\bibinfo {year} {2020})},\ \Eprint
  {http://arxiv.org/abs/1911.09855} {arXiv:1911.09855 [gr-qc]} \BibitemShut
  {NoStop}%
\bibitem [{\citenamefont {Gan}\ \emph {et~al.}(2019{\natexlab{b}})\citenamefont
  {Gan}, \citenamefont {Wang}, \citenamefont {Wu},\ and\ \citenamefont
  {Yang}}]{Gan:2019ibg}%
  \BibitemOpen
  \bibfield  {author} {\bibinfo {author} {\bibfnamefont {Q.}~\bibnamefont
  {Gan}}, \bibinfo {author} {\bibfnamefont {P.}~\bibnamefont {Wang}}, \bibinfo
  {author} {\bibfnamefont {H.}~\bibnamefont {Wu}}, \ and\ \bibinfo {author}
  {\bibfnamefont {H.}~\bibnamefont {Yang}},\ }\href@noop {} {\  (\bibinfo
  {year} {2019}{\natexlab{b}})},\ \Eprint {http://arxiv.org/abs/1911.10996}
  {arXiv:1911.10996 [gr-qc]} \BibitemShut {NoStop}%
\bibitem [{\citenamefont {Hollands}\ \emph {et~al.}(2019)\citenamefont
  {Hollands}, \citenamefont {Wald},\ and\ \citenamefont
  {Zahn}}]{Hollands:2019whz}%
  \BibitemOpen
  \bibfield  {author} {\bibinfo {author} {\bibfnamefont {S.}~\bibnamefont
  {Hollands}}, \bibinfo {author} {\bibfnamefont {R.~M.}\ \bibnamefont {Wald}},
  \ and\ \bibinfo {author} {\bibfnamefont {J.}~\bibnamefont {Zahn}},\
  }\href@noop {} {\  (\bibinfo {year} {2019})},\ \Eprint
  {http://arxiv.org/abs/1912.06047} {arXiv:1912.06047 [gr-qc]} \BibitemShut
  {NoStop}%
\bibitem [{\citenamefont {Rahman}\ \emph {et~al.}(2020)\citenamefont {Rahman},
  \citenamefont {Mitra},\ and\ \citenamefont {Chakraborty}}]{Rahman:2020guv}%
  \BibitemOpen
  \bibfield  {author} {\bibinfo {author} {\bibfnamefont {M.}~\bibnamefont
  {Rahman}}, \bibinfo {author} {\bibfnamefont {S.}~\bibnamefont {Mitra}}, \
  and\ \bibinfo {author} {\bibfnamefont {S.}~\bibnamefont {Chakraborty}},\
  }\href@noop {} {\  (\bibinfo {year} {2020})},\ \Eprint
  {http://arxiv.org/abs/2001.00599} {arXiv:2001.00599 [gr-qc]} \BibitemShut
  {NoStop}%
\bibitem [{\citenamefont {Mishra}(2020)}]{Mishra:2020gce}%
  \BibitemOpen
  \bibfield  {author} {\bibinfo {author} {\bibfnamefont {A.~K.}\ \bibnamefont
  {Mishra}},\ }\href@noop {} {\  (\bibinfo {year} {2020})},\ \Eprint
  {http://arxiv.org/abs/2004.01243} {arXiv:2004.01243 [gr-qc]} \BibitemShut
  {NoStop}%
\bibitem [{\citenamefont {Emparan}\ and\ \citenamefont
  {Toma\v{s}evi\'c}(2020)}]{Emparan:2020rnp}%
  \BibitemOpen
  \bibfield  {author} {\bibinfo {author} {\bibfnamefont {R.}~\bibnamefont
  {Emparan}}\ and\ \bibinfo {author} {\bibfnamefont {M.}~\bibnamefont
  {Toma\v{s}evi\'c}},\ }\href@noop {} {\  (\bibinfo {year} {2020})},\ \Eprint
  {http://arxiv.org/abs/2002.02083} {arXiv:2002.02083 [hep-th]} \BibitemShut
  {NoStop}%
\bibitem [{\citenamefont {Casals}\ and\ \citenamefont
  {Marinho}(2020)}]{casals2020glimpses}%
  \BibitemOpen
  \bibfield  {author} {\bibinfo {author} {\bibfnamefont {M.}~\bibnamefont
  {Casals}}\ and\ \bibinfo {author} {\bibfnamefont {C.~I.~S.}\ \bibnamefont
  {Marinho}},\ }\href@noop {} {\enquote {\bibinfo {title} {Glimpses of
  violation of strong cosmic censorship in rotating black holes},}\ } (\bibinfo
  {year} {2020}),\ \Eprint {http://arxiv.org/abs/2006.06483} {arXiv:2006.06483
  [gr-qc]} \BibitemShut {NoStop}%
\bibitem [{\citenamefont {Weyl}(1917)}]{Weyl}%
  \BibitemOpen
  \bibfield  {author} {\bibinfo {author} {\bibfnamefont {H.}~\bibnamefont
  {Weyl}},\ }\href {\doibase 10.1002/andp.19173591804} {\bibfield  {journal}
  {\bibinfo  {journal} {Annalen der Physik}\ }\textbf {\bibinfo {volume}
  {359}},\ \bibinfo {pages} {117} (\bibinfo {year} {1917})},\ \Eprint
  {http://arxiv.org/abs/https://onlinelibrary.wiley.com/doi/pdf/10.1002/andp.19173591804}
  {https://onlinelibrary.wiley.com/doi/pdf/10.1002/andp.19173591804}
  \BibitemShut {NoStop}%
\bibitem [{\citenamefont {Griffiths}\ and\ \citenamefont
  {Podolský}(2009)}]{griffiths_podolsky_2009}%
  \BibitemOpen
  \bibfield  {author} {\bibinfo {author} {\bibfnamefont {J.~B.}\ \bibnamefont
  {Griffiths}}\ and\ \bibinfo {author} {\bibfnamefont {J.}~\bibnamefont
  {Podolský}},\ }\href {\doibase 10.1017/CBO9780511635397} {\emph {\bibinfo
  {title} {Exact Space-Times in Einstein's General Relativity}}},\ Cambridge
  Monographs on Mathematical Physics\ (\bibinfo  {publisher} {Cambridge
  University Press},\ \bibinfo {year} {2009})\BibitemShut {NoStop}%
\bibitem [{\citenamefont {Appels}\ \emph {et~al.}(2016)\citenamefont {Appels},
  \citenamefont {Gregory},\ and\ \citenamefont {Kubiz\ifmmode~\check{n}\else
  \v{n}\fi{}\'ak}}]{Gregory-PRL-2016}%
  \BibitemOpen
  \bibfield  {author} {\bibinfo {author} {\bibfnamefont {M.}~\bibnamefont
  {Appels}}, \bibinfo {author} {\bibfnamefont {R.}~\bibnamefont {Gregory}}, \
  and\ \bibinfo {author} {\bibfnamefont {D.}~\bibnamefont
  {Kubiz\ifmmode~\check{n}\else \v{n}\fi{}\'ak}},\ }\href {\doibase
  10.1103/PhysRevLett.117.131303} {\bibfield  {journal} {\bibinfo  {journal}
  {Phys. Rev. Lett.}\ }\textbf {\bibinfo {volume} {117}},\ \bibinfo {pages}
  {131303} (\bibinfo {year} {2016})}\BibitemShut {NoStop}%
\bibitem [{\citenamefont {Gregory}\ and\ \citenamefont
  {Scoins}(2019)}]{GREGORY2019191}%
  \BibitemOpen
  \bibfield  {author} {\bibinfo {author} {\bibfnamefont {R.}~\bibnamefont
  {Gregory}}\ and\ \bibinfo {author} {\bibfnamefont {A.}~\bibnamefont
  {Scoins}},\ }\href@noop {} {\bibfield  {journal} {\bibinfo  {journal}
  {Physics Letters B}\ }\textbf {\bibinfo {volume} {796}},\ \bibinfo {pages}
  {191} (\bibinfo {year} {2019})}\BibitemShut {NoStop}%
\bibitem [{\citenamefont {Anabalón}\ \emph {et~al.}(2019)\citenamefont
  {Anabalón}, \citenamefont {Gray}, \citenamefont {Gregory}, \citenamefont
  {Kubizňák},\ and\ \citenamefont {Mann}}]{Anabalon:2018qfv}%
  \BibitemOpen
  \bibfield  {author} {\bibinfo {author} {\bibfnamefont {A.}~\bibnamefont
  {Anabalón}}, \bibinfo {author} {\bibfnamefont {F.}~\bibnamefont {Gray}},
  \bibinfo {author} {\bibfnamefont {R.}~\bibnamefont {Gregory}}, \bibinfo
  {author} {\bibfnamefont {D.}~\bibnamefont {Kubizňák}}, \ and\ \bibinfo
  {author} {\bibfnamefont {R.~B.}\ \bibnamefont {Mann}},\ }\href {\doibase
  10.1007/JHEP04(2019)096} {\bibfield  {journal} {\bibinfo  {journal} {JHEP}\
  }\textbf {\bibinfo {volume} {04}},\ \bibinfo {pages} {096} (\bibinfo {year}
  {2019})},\ \Eprint {http://arxiv.org/abs/1811.04936} {arXiv:1811.04936
  [hep-th]} \BibitemShut {NoStop}%
\bibitem [{\citenamefont {Anabalón}\ \emph {et~al.}(2018)\citenamefont
  {Anabalón}, \citenamefont {Appels}, \citenamefont {Gregory}, \citenamefont
  {Kubizňák}, \citenamefont {Mann},\ and\ \citenamefont
  {Ovgün}}]{Anabalon:2018ydc}%
  \BibitemOpen
  \bibfield  {author} {\bibinfo {author} {\bibfnamefont {A.}~\bibnamefont
  {Anabalón}}, \bibinfo {author} {\bibfnamefont {M.}~\bibnamefont {Appels}},
  \bibinfo {author} {\bibfnamefont {R.}~\bibnamefont {Gregory}}, \bibinfo
  {author} {\bibfnamefont {D.}~\bibnamefont {Kubizňák}}, \bibinfo {author}
  {\bibfnamefont {R.~B.}\ \bibnamefont {Mann}}, \ and\ \bibinfo {author}
  {\bibfnamefont {A.}~\bibnamefont {Ovgün}},\ }\href {\doibase
  10.1103/PhysRevD.98.104038} {\bibfield  {journal} {\bibinfo  {journal} {Phys.
  Rev.}\ }\textbf {\bibinfo {volume} {D98}},\ \bibinfo {pages} {104038}
  (\bibinfo {year} {2018})},\ \Eprint {http://arxiv.org/abs/1805.02687}
  {arXiv:1805.02687 [hep-th]} \BibitemShut {NoStop}%
\bibitem [{\citenamefont {Ashtekar}\ and\ \citenamefont
  {Dray}(1981)}]{ashtekar1981}%
  \BibitemOpen
  \bibfield  {author} {\bibinfo {author} {\bibfnamefont {A.}~\bibnamefont
  {Ashtekar}}\ and\ \bibinfo {author} {\bibfnamefont {T.}~\bibnamefont
  {Dray}},\ }\href {https://projecteuclid.org:443/euclid.cmp/1103909143}
  {\bibfield  {journal} {\bibinfo  {journal} {Comm. Math. Phys.}\ }\textbf
  {\bibinfo {volume} {79}},\ \bibinfo {pages} {581} (\bibinfo {year}
  {1981})}\BibitemShut {NoStop}%
\bibitem [{\citenamefont {Podolsky}\ \emph {et~al.}()\citenamefont {Podolsky},
  \citenamefont {Ortaggio},\ and\ \citenamefont {Krtouss}}]{Podolsky-2003}%
  \BibitemOpen
  \bibfield  {author} {\bibinfo {author} {\bibfnamefont {J.}~\bibnamefont
  {Podolsky}}, \bibinfo {author} {\bibfnamefont {M.}~\bibnamefont {Ortaggio}},
  \ and\ \bibinfo {author} {\bibfnamefont {P.}~\bibnamefont {Krtouss}},\
  }\href@noop {} {\bibfield  {journal} {\bibinfo  {journal} {Phys. Rev. D}\
  }\textbf {\bibinfo {volume} {68}},\ \bibinfo {pages} {124004}}\BibitemShut
  {NoStop}%
\bibitem [{\citenamefont {Bicak}\ \emph {et~al.}()\citenamefont {Bicak},
  \citenamefont {Reilly},\ and\ \citenamefont {Winicour}}]{Bicak-Winicour}%
  \BibitemOpen
  \bibfield  {author} {\bibinfo {author} {\bibfnamefont {J.}~\bibnamefont
  {Bicak}}, \bibinfo {author} {\bibfnamefont {P.}~\bibnamefont {Reilly}}, \
  and\ \bibinfo {author} {\bibfnamefont {J.}~\bibnamefont {Winicour}},\
  }\href@noop {} {\bibfield  {journal} {\bibinfo  {journal} {Gen. Relat.
  Grav.}\ }\textbf {\bibinfo {volume} {20}}}\BibitemShut {NoStop}%
\bibitem [{\citenamefont {R.~Gómez}\ and\ \citenamefont
  {Winicour}()}]{Winicour}%
  \BibitemOpen
  \bibfield  {author} {\bibinfo {author} {\bibfnamefont {P.~P.}\ \bibnamefont
  {R.~Gómez}}\ and\ \bibinfo {author} {\bibfnamefont {J.}~\bibnamefont
  {Winicour}},\ }\href@noop {} {\bibfield  {journal} {\bibinfo  {journal}
  {Journal of Mathematical Physics}\ }\textbf {\bibinfo {volume}
  {35}}}\BibitemShut {NoStop}%
\bibitem [{\citenamefont {Hawking}\ and\ \citenamefont
  {Ross}(1995)}]{Hawking-Ross-PRL-1995}%
  \BibitemOpen
  \bibfield  {author} {\bibinfo {author} {\bibfnamefont {S.~W.}\ \bibnamefont
  {Hawking}}\ and\ \bibinfo {author} {\bibfnamefont {S.~F.}\ \bibnamefont
  {Ross}},\ }\href {\doibase 10.1103/PhysRevLett.75.3382} {\bibfield  {journal}
  {\bibinfo  {journal} {Phys. Rev. Lett.}\ }\textbf {\bibinfo {volume} {75}},\
  \bibinfo {pages} {3382} (\bibinfo {year} {1995})}\BibitemShut {NoStop}%
\bibitem [{\citenamefont {Hawking}\ and\ \citenamefont
  {Ross}(1997)}]{Hawking:1997ia}%
  \BibitemOpen
  \bibfield  {author} {\bibinfo {author} {\bibfnamefont {S.~W.}\ \bibnamefont
  {Hawking}}\ and\ \bibinfo {author} {\bibfnamefont {S.~F.}\ \bibnamefont
  {Ross}},\ }\href {\doibase 10.1103/PhysRevD.56.6403} {\bibfield  {journal}
  {\bibinfo  {journal} {Phys. Rev.}\ }\textbf {\bibinfo {volume} {D56}},\
  \bibinfo {pages} {6403} (\bibinfo {year} {1997})},\ \Eprint
  {http://arxiv.org/abs/hep-th/9705147} {arXiv:hep-th/9705147 [hep-th]}
  \BibitemShut {NoStop}%
\bibitem [{\citenamefont {Eardley}\ \emph {et~al.}(1995)\citenamefont
  {Eardley}, \citenamefont {Horowitz}, \citenamefont {Kastor},\ and\
  \citenamefont {Traschen}}]{Earley-PRL-1995}%
  \BibitemOpen
  \bibfield  {author} {\bibinfo {author} {\bibfnamefont {D.~M.}\ \bibnamefont
  {Eardley}}, \bibinfo {author} {\bibfnamefont {G.~T.}\ \bibnamefont
  {Horowitz}}, \bibinfo {author} {\bibfnamefont {D.~A.}\ \bibnamefont
  {Kastor}}, \ and\ \bibinfo {author} {\bibfnamefont {J.}~\bibnamefont
  {Traschen}},\ }\href {\doibase 10.1103/PhysRevLett.75.3390} {\bibfield
  {journal} {\bibinfo  {journal} {Phys. Rev. Lett.}\ }\textbf {\bibinfo
  {volume} {75}},\ \bibinfo {pages} {3390} (\bibinfo {year}
  {1995})}\BibitemShut {NoStop}%
\bibitem [{\citenamefont {Emparan}\ and\ \citenamefont
  {Reall}(2002)}]{Emparan-Reall-PRL-2002}%
  \BibitemOpen
  \bibfield  {author} {\bibinfo {author} {\bibfnamefont {R.}~\bibnamefont
  {Emparan}}\ and\ \bibinfo {author} {\bibfnamefont {H.~S.}\ \bibnamefont
  {Reall}},\ }\href {\doibase 10.1103/PhysRevLett.88.101101} {\bibfield
  {journal} {\bibinfo  {journal} {Phys. Rev. Lett.}\ }\textbf {\bibinfo
  {volume} {88}},\ \bibinfo {pages} {101101} (\bibinfo {year}
  {2002})}\BibitemShut {NoStop}%
\bibitem [{\citenamefont {Griffiths}\ \emph {et~al.}(2006)\citenamefont
  {Griffiths}, \citenamefont {Krtous},\ and\ \citenamefont
  {Podolsky}}]{Griffiths:2006tk}%
  \BibitemOpen
  \bibfield  {author} {\bibinfo {author} {\bibfnamefont {J.~B.}\ \bibnamefont
  {Griffiths}}, \bibinfo {author} {\bibfnamefont {P.}~\bibnamefont {Krtous}}, \
  and\ \bibinfo {author} {\bibfnamefont {J.}~\bibnamefont {Podolsky}},\ }\href
  {\doibase 10.1088/0264-9381/23/23/008} {\bibfield  {journal} {\bibinfo
  {journal} {Class. Quant. Grav.}\ }\textbf {\bibinfo {volume} {23}},\ \bibinfo
  {pages} {6745} (\bibinfo {year} {2006})},\ \Eprint
  {http://arxiv.org/abs/gr-qc/0609056} {arXiv:gr-qc/0609056 [gr-qc]}
  \BibitemShut {NoStop}%
\bibitem [{\citenamefont {Horowitz}\ and\ \citenamefont
  {Sheinblatt}(1997)}]{Horowitz:1996yb}%
  \BibitemOpen
  \bibfield  {author} {\bibinfo {author} {\bibfnamefont {G.~T.}\ \bibnamefont
  {Horowitz}}\ and\ \bibinfo {author} {\bibfnamefont {H.~J.}\ \bibnamefont
  {Sheinblatt}},\ }\href {\doibase 10.1103/PhysRevD.55.650} {\bibfield
  {journal} {\bibinfo  {journal} {Phys. Rev. D}\ }\textbf {\bibinfo {volume}
  {55}},\ \bibinfo {pages} {650} (\bibinfo {year} {1997})},\ \Eprint
  {http://arxiv.org/abs/gr-qc/9607027} {arXiv:gr-qc/9607027} \BibitemShut
  {NoStop}%
\bibitem [{\citenamefont {Destounis}\ \emph {et~al.}(2020)\citenamefont
  {Destounis}, \citenamefont {Fontana},\ and\ \citenamefont
  {Mena}}]{Destounis:2020pjk}%
  \BibitemOpen
  \bibfield  {author} {\bibinfo {author} {\bibfnamefont {K.}~\bibnamefont
  {Destounis}}, \bibinfo {author} {\bibfnamefont {R.~D.}\ \bibnamefont
  {Fontana}}, \ and\ \bibinfo {author} {\bibfnamefont {F.~C.}\ \bibnamefont
  {Mena}},\ }\href@noop {} {\  (\bibinfo {year} {2020})},\ \Eprint
  {http://arxiv.org/abs/2005.03028} {arXiv:2005.03028 [gr-qc]} \BibitemShut
  {NoStop}%
\bibitem [{\citenamefont {Wells}(1998)}]{Wells}%
  \BibitemOpen
  \bibfield  {author} {\bibinfo {author} {\bibfnamefont {C.~G.}\ \bibnamefont
  {Wells}},\ }\href@noop {} {\  (\bibinfo {year} {1998})},\ \Eprint
  {http://arxiv.org/abs/9808044} {arXiv:9808044 [gr-qc]} \BibitemShut {NoStop}%
\bibitem [{\citenamefont {Plebanski}\ and\ \citenamefont
  {Demianski}(1976)}]{Plebanski:1976gy}%
  \BibitemOpen
  \bibfield  {author} {\bibinfo {author} {\bibfnamefont {J.~F.}\ \bibnamefont
  {Plebanski}}\ and\ \bibinfo {author} {\bibfnamefont {M.}~\bibnamefont
  {Demianski}},\ }\href {\doibase 10.1016/0003-4916(76)90240-2} {\bibfield
  {journal} {\bibinfo  {journal} {Annals Phys.}\ }\textbf {\bibinfo {volume}
  {98}},\ \bibinfo {pages} {98} (\bibinfo {year} {1976})}\BibitemShut {NoStop}%
\bibitem [{\citenamefont {Gregory}(2017)}]{Gregory:2017ogk}%
  \BibitemOpen
  \bibfield  {author} {\bibinfo {author} {\bibfnamefont {R.}~\bibnamefont
  {Gregory}},\ }\href {\doibase 10.1088/1742-6596/942/1/012002} {\bibfield
  {journal} {\bibinfo  {journal} {J. Phys. Conf. Ser.}\ }\textbf {\bibinfo
  {volume} {942}},\ \bibinfo {pages} {012002} (\bibinfo {year} {2017})},\
  \Eprint {http://arxiv.org/abs/1712.04992} {arXiv:1712.04992 [hep-th]}
  \BibitemShut {NoStop}%
\bibitem [{\citenamefont {Kinnersley}\ and\ \citenamefont
  {Walker}(1970)}]{Kinnersley_1970}%
  \BibitemOpen
  \bibfield  {author} {\bibinfo {author} {\bibfnamefont {W.}~\bibnamefont
  {Kinnersley}}\ and\ \bibinfo {author} {\bibfnamefont {M.}~\bibnamefont
  {Walker}},\ }\href {\doibase 10.1103/PhysRevD.2.1359} {\bibfield  {journal}
  {\bibinfo  {journal} {Phys. Rev. D}\ }\textbf {\bibinfo {volume} {2}},\
  \bibinfo {pages} {1359} (\bibinfo {year} {1970})}\BibitemShut {NoStop}%
\bibitem [{\citenamefont {Wald}(1984)}]{Wald-book}%
  \BibitemOpen
  \bibfield  {author} {\bibinfo {author} {\bibfnamefont {R.~M.}\ \bibnamefont
  {Wald}},\ }\href {\doibase 10.7208/chicago/9780226870373.001.0001} {\emph
  {\bibinfo {title} {General Relativity}}},\ University of Chicago Press\
  (\bibinfo  {publisher} {University of Chicago Press},\ \bibinfo {year}
  {1984})\BibitemShut {NoStop}%
\bibitem [{\citenamefont {Bini}\ \emph {et~al.}(2008)\citenamefont {Bini},
  \citenamefont {Cherubini},\ and\ \citenamefont {Geralico}}]{Bini:2014kga}%
  \BibitemOpen
  \bibfield  {author} {\bibinfo {author} {\bibfnamefont {D.}~\bibnamefont
  {Bini}}, \bibinfo {author} {\bibfnamefont {C.}~\bibnamefont {Cherubini}}, \
  and\ \bibinfo {author} {\bibfnamefont {A.}~\bibnamefont {Geralico}},\ }\href
  {\doibase 10.1063/1.2938699} {\bibfield  {journal} {\bibinfo  {journal} {J.
  Math. Phys.}\ }\textbf {\bibinfo {volume} {49}},\ \bibinfo {pages} {062502}
  (\bibinfo {year} {2008})},\ \Eprint {http://arxiv.org/abs/1408.4593}
  {arXiv:1408.4593 [gr-qc]} \BibitemShut {NoStop}%
\bibitem [{\citenamefont {Charmousis}\ \emph {et~al.}(2009)\citenamefont
  {Charmousis}, \citenamefont {Kolyvaris},\ and\ \citenamefont
  {Papantonopoulos}}]{Charmousis_2009}%
  \BibitemOpen
  \bibfield  {author} {\bibinfo {author} {\bibfnamefont {C.}~\bibnamefont
  {Charmousis}}, \bibinfo {author} {\bibfnamefont {T.}~\bibnamefont
  {Kolyvaris}}, \ and\ \bibinfo {author} {\bibfnamefont {E.}~\bibnamefont
  {Papantonopoulos}},\ }\href {\doibase 10.1088/0264-9381/26/17/175012}
  {\bibfield  {journal} {\bibinfo  {journal} {Classical and Quantum Gravity}\
  }\textbf {\bibinfo {volume} {26}},\ \bibinfo {pages} {175012} (\bibinfo
  {year} {2009})}\BibitemShut {NoStop}%
\bibitem [{\citenamefont {Jansen}(2017)}]{Jansen:2017oag}%
  \BibitemOpen
  \bibfield  {author} {\bibinfo {author} {\bibfnamefont {A.}~\bibnamefont
  {Jansen}},\ }\href {\doibase 10.1140/epjp/i2017-11825-9} {\bibfield
  {journal} {\bibinfo  {journal} {Eur. Phys. J. Plus}\ }\textbf {\bibinfo
  {volume} {132}},\ \bibinfo {pages} {546} (\bibinfo {year} {2017})},\ \Eprint
  {http://arxiv.org/abs/1709.09178} {arXiv:1709.09178 [gr-qc]} \BibitemShut
  {NoStop}%
\bibitem [{\citenamefont {Dias}\ \emph {et~al.}(2016)\citenamefont {Dias},
  \citenamefont {Santos},\ and\ \citenamefont {Way}}]{Dias:2015nua}%
  \BibitemOpen
  \bibfield  {author} {\bibinfo {author} {\bibfnamefont {O.~J.~C.}\
  \bibnamefont {Dias}}, \bibinfo {author} {\bibfnamefont {J.~E.}\ \bibnamefont
  {Santos}}, \ and\ \bibinfo {author} {\bibfnamefont {B.}~\bibnamefont {Way}},\
  }\href {\doibase 10.1088/0264-9381/33/13/133001} {\bibfield  {journal}
  {\bibinfo  {journal} {Class. Quant. Grav.}\ }\textbf {\bibinfo {volume}
  {33}},\ \bibinfo {pages} {133001} (\bibinfo {year} {2016})},\ \Eprint
  {http://arxiv.org/abs/1510.02804} {arXiv:1510.02804 [hep-th]} \BibitemShut
  {NoStop}%
\bibitem [{\citenamefont {Berti}\ \emph {et~al.}(2007)\citenamefont {Berti},
  \citenamefont {Cardoso}, \citenamefont {Gonzalez},\ and\ \citenamefont
  {Sperhake}}]{Berti:2007dg}%
  \BibitemOpen
  \bibfield  {author} {\bibinfo {author} {\bibfnamefont {E.}~\bibnamefont
  {Berti}}, \bibinfo {author} {\bibfnamefont {V.}~\bibnamefont {Cardoso}},
  \bibinfo {author} {\bibfnamefont {J.~A.}\ \bibnamefont {Gonzalez}}, \ and\
  \bibinfo {author} {\bibfnamefont {U.}~\bibnamefont {Sperhake}},\ }\href
  {\doibase 10.1103/PhysRevD.75.124017} {\bibfield  {journal} {\bibinfo
  {journal} {Phys. Rev.}\ }\textbf {\bibinfo {volume} {D75}},\ \bibinfo {pages}
  {124017} (\bibinfo {year} {2007})},\ \Eprint
  {http://arxiv.org/abs/gr-qc/0701086} {arXiv:gr-qc/0701086 [gr-qc]}
  \BibitemShut {NoStop}%
\bibitem [{\citenamefont {Destounis}(2019{\natexlab{b}})}]{Destounis:2019hca}%
  \BibitemOpen
  \bibfield  {author} {\bibinfo {author} {\bibfnamefont {K.}~\bibnamefont
  {Destounis}},\ }\href {\doibase 10.1103/PhysRevD.100.044054} {\bibfield
  {journal} {\bibinfo  {journal} {Phys. Rev. D}\ }\textbf {\bibinfo {volume}
  {100}},\ \bibinfo {pages} {044054} (\bibinfo {year} {2019}{\natexlab{b}})},\
  \Eprint {http://arxiv.org/abs/1908.06117} {arXiv:1908.06117 [gr-qc]}
  \BibitemShut {NoStop}%
\bibitem [{\citenamefont {Hod}(2017)}]{Hod:2017gvn}%
  \BibitemOpen
  \bibfield  {author} {\bibinfo {author} {\bibfnamefont {S.}~\bibnamefont
  {Hod}},\ }\href {\doibase 10.1140/epjc/s10052-017-4920-8} {\bibfield
  {journal} {\bibinfo  {journal} {Eur. Phys. J.}\ }\textbf {\bibinfo {volume}
  {C77}},\ \bibinfo {pages} {351} (\bibinfo {year} {2017})},\ \Eprint
  {http://arxiv.org/abs/1705.04726} {arXiv:1705.04726 [hep-th]} \BibitemShut
  {NoStop}%
\bibitem [{\citenamefont {Christodoulou}(2008)}]{Christodoulou:2008nj}%
  \BibitemOpen
  \bibfield  {author} {\bibinfo {author} {\bibfnamefont {D.}~\bibnamefont
  {Christodoulou}}\ }(\bibinfo {year} {2008})\ pp.\ \bibinfo {pages} {24--34},\
  \Eprint {http://arxiv.org/abs/0805.3880} {arXiv:0805.3880 [gr-qc]}
  \BibitemShut {NoStop}%
\bibitem [{\citenamefont {Brady}\ \emph {et~al.}(1998)\citenamefont {Brady},
  \citenamefont {Moss},\ and\ \citenamefont {Myers}}]{Brady:1998au}%
  \BibitemOpen
  \bibfield  {author} {\bibinfo {author} {\bibfnamefont {P.~R.}\ \bibnamefont
  {Brady}}, \bibinfo {author} {\bibfnamefont {I.~G.}\ \bibnamefont {Moss}}, \
  and\ \bibinfo {author} {\bibfnamefont {R.~C.}\ \bibnamefont {Myers}},\ }\href
  {\doibase 10.1103/PhysRevLett.80.3432} {\bibfield  {journal} {\bibinfo
  {journal} {Phys. Rev. Lett.}\ }\textbf {\bibinfo {volume} {80}},\ \bibinfo
  {pages} {3432} (\bibinfo {year} {1998})},\ \Eprint
  {http://arxiv.org/abs/gr-qc/9801032} {arXiv:gr-qc/9801032 [gr-qc]}
  \BibitemShut {NoStop}%
\bibitem [{\citenamefont {Gibbons}\ \emph {et~al.}(1990)\citenamefont
  {Gibbons}, \citenamefont {Hawking},\ and\ \citenamefont
  {Vachaspati}}]{Gibbons:1990gp}%
  \BibitemOpen
  \bibinfo {editor} {\bibfnamefont {G.}~\bibnamefont {Gibbons}}, \bibinfo
  {editor} {\bibfnamefont {S.}~\bibnamefont {Hawking}}, \ and\ \bibinfo
  {editor} {\bibfnamefont {T.}~\bibnamefont {Vachaspati}},\ eds.,\ \href@noop
  {} {\emph {\bibinfo {title} {{The Formation and evolution of cosmic strings.
  Proceedings, Workshop, Cambridge, UK, July 3-7, 1989}}}}\ (\bibinfo {year}
  {1990})\BibitemShut {NoStop}%
\end{thebibliography}%
\end{document}